\documentclass[aps,prl,preprint,tightenlines,superscriptaddress,showpacs,byrevtex]{revtex4}

\usepackage{epsfig,floatflt}
\usepackage{graphicx} % Include figure files
\usepackage{dcolumn}  % Align table columns on decimal point
\usepackage{amssymb,amsmath,indentfirst,enumerate}

\graphicspath{{ps}}

\renewcommand{\arraystretch}{1.1}

\newcommand{\UFS}{\ensuremath{\Upsilon(10860)}}
\newcommand{\Uf}{\ensuremath{\Upsilon(10860)}}
\newcommand{\Un}{\ensuremath{\Upsilon(nS)}}
\newcommand{\Uo}{\ensuremath{\Upsilon(1S)}}
\newcommand{\Ut}{\ensuremath{\Upsilon(2S)}}
\newcommand{\Uh}{\ensuremath{\Upsilon(3S)}}
\newcommand{\hm}{\ensuremath{h_b(mP)}}
\newcommand{\hb}{\ensuremath{h_b(1P)}}
\newcommand{\hp}{\ensuremath{h_b(2P)}}
\newcommand{\pp}{\ensuremath{\pi^+\pi^-}}
\newcommand{\uu}{\ensuremath{\mu^+\mu^-}}
\newcommand{\ee}{\ensuremath{e^+e^-}}
\newcommand{\Zbl}{\ensuremath{Z_b(10610)}}
\newcommand{\Zbh}{\ensuremath{Z_b(10650)}}

\newcommand{\mmpp}{M_r(\pi^+\pi^-)}
\newcommand{\mmp}{M_r(\pi)}

\newcommand{\gev}{\,\mathrm{GeV}}

\newcommand{\gevm}{\mathrm{GeV}/c^2}

\newcommand{\mevm}{\mathrm{MeV}/c^2}

\newcommand{\bbpi}{\ensuremath{B^{(*)}B^{(*)}\pi}}
\newcommand{\bbstpi}{\ensuremath{BB^*\pi}}
\newcommand{\bstbstpi}{\ensuremath{B^{*}B^{*}\pi}}
\def\nima#1#2#3{{Nucl.\ Instr.\ and Meth.} {\bf A#1}, #3 (#2)}

\def\prd#1#2#3{{ Phys.\ Rev.}   {\bf D#1}, #3 (#2)}

\def\prl#1#2#3{{ Phys.\ Rev.\ Lett.} {\bf #1}, #3 (#2)}

\begin{document}

\preprint{\vbox{ \hbox{   }
                 \hbox{BELLE-CONF-1272}
%                 \hbox{hep-ex nnnn, if available}
}}

\title{\quad\\[0.5cm]
Study of Three-Body {\boldmath $\UFS$} Decays}

%%% Paper:
%%% Journal:  2012 Conference Papers
%%% March 12, 2012 - first draft
%%% April 13, 2012 - second draft: add newcomers (BH Kim, B Kronenbitter)
%%%                  and update affiliations (D Mohapatra, N Muramatsu)
%%% May 8, 2012 - third draft: add V Chekelian, J Klucar, S Sandilya, L Santelj
%%% June 17, 2012 - fourth draft: remove H Ha, S H Kim
%%% August 14, 2012 - fifth draft: add Bilbao (2 authors), Goettingen (4 authors)
%%% Non-responding authors or those who said NO are commented out.
%%% ====================================================================
%%% Click the RELOAD button on your web browser to see the updated file.
%%% ====================================================================
%%% Use \input{author} to insert this material into your latex file.
%%%%% Force institutions to appear in alphabetical order when typeset.
\affiliation{University of the Basque Country UPV/EHU, Bilbao}
\affiliation{University of Bonn, Bonn}
\affiliation{Budker Institute of Nuclear Physics SB RAS and Novosibirsk State University, Novosibirsk 630090}
\affiliation{Faculty of Mathematics and Physics, Charles University, Prague}
\affiliation{Chiba University, Chiba}
\affiliation{University of Cincinnati, Cincinnati, Ohio 45221}
\affiliation{Department of Physics, Fu Jen Catholic University, Taipei}
\affiliation{Justus-Liebig-Universit\"at Gie\ss{}en, Gie\ss{}en}
\affiliation{Gifu University, Gifu}
\affiliation{II. Physikalisches Institut, Georg-August-Universit\"at G\"ottingen, G\"ottingen}
\affiliation{The Graduate University for Advanced Studies, Hayama}
\affiliation{Gyeongsang National University, Chinju}
\affiliation{Hanyang University, Seoul}
\affiliation{University of Hawaii, Honolulu, Hawaii 96822}
\affiliation{High Energy Accelerator Research Organization (KEK), Tsukuba}
\affiliation{Hiroshima Institute of Technology, Hiroshima}
\affiliation{IKERBASQUE, Bilbao}
\affiliation{University of Illinois at Urbana-Champaign, Urbana, Illinois 61801}
\affiliation{Indian Institute of Technology Guwahati, Guwahati}
\affiliation{Indian Institute of Technology Madras, Madras}
\affiliation{Indiana University, Bloomington, Indiana 47408}
\affiliation{Institute of High Energy Physics, Chinese Academy of Sciences, Beijing}
\affiliation{Institute of High Energy Physics, Vienna}
\affiliation{Institute of High Energy Physics, Protvino}
\affiliation{Institute of Mathematical Sciences, Chennai}
\affiliation{INFN - Sezione di Torino, Torino}
\affiliation{Institute for Theoretical and Experimental Physics, Moscow}
\affiliation{J. Stefan Institute, Ljubljana}
\affiliation{Kanagawa University, Yokohama}
\affiliation{Institut f\"ur Experimentelle Kernphysik, Karlsruher Institut f\"ur Technologie, Karlsruhe}
\affiliation{Korea Institute of Science and Technology Information, Daejeon}
\affiliation{Korea University, Seoul}
\affiliation{Kyoto University, Kyoto}
\affiliation{Kyungpook National University, Taegu}
\affiliation{\'Ecole Polytechnique F\'ed\'erale de Lausanne (EPFL), Lausanne}
\affiliation{Faculty of Mathematics and Physics, University of Ljubljana, Ljubljana}
\affiliation{Luther College, Decorah, Iowa 52101}
\affiliation{University of Maribor, Maribor}
\affiliation{Max-Planck-Institut f\"ur Physik, M\"unchen}
\affiliation{University of Melbourne, School of Physics, Victoria 3010}
\affiliation{Graduate School of Science, Nagoya University, Nagoya}
\affiliation{Kobayashi-Maskawa Institute, Nagoya University, Nagoya}
\affiliation{Nara University of Education, Nara}
\affiliation{Nara Women's University, Nara}
\affiliation{National Central University, Chung-li}
\affiliation{National United University, Miao Li}
\affiliation{Department of Physics, National Taiwan University, Taipei}
\affiliation{H. Niewodniczanski Institute of Nuclear Physics, Krakow}
\affiliation{Nippon Dental University, Niigata}
\affiliation{Niigata University, Niigata}
\affiliation{University of Nova Gorica, Nova Gorica}
\affiliation{Osaka City University, Osaka}
\affiliation{Osaka University, Osaka}
\affiliation{Pacific Northwest National Laboratory, Richland, Washington 99352}
\affiliation{Panjab University, Chandigarh}
\affiliation{Peking University, Beijing}
\affiliation{Princeton University, Princeton, New Jersey 08544}
\affiliation{Research Center for Electron Photon Science, Tohoku University, Sendai}
\affiliation{Research Center for Nuclear Physics, Osaka University, Osaka}
\affiliation{RIKEN BNL Research Center, Upton, New York 11973}
\affiliation{Saga University, Saga}
\affiliation{University of Science and Technology of China, Hefei}
\affiliation{Seoul National University, Seoul}
\affiliation{Shinshu University, Nagano}
\affiliation{Sungkyunkwan University, Suwon}
\affiliation{School of Physics, University of Sydney, NSW 2006}
\affiliation{Tata Institute of Fundamental Research, Mumbai}
\affiliation{Excellence Cluster Universe, Technische Universit\"at M\"unchen, Garching}
\affiliation{Toho University, Funabashi}
\affiliation{Tohoku Gakuin University, Tagajo}
\affiliation{Tohoku University, Sendai}
\affiliation{Department of Physics, University of Tokyo, Tokyo}
\affiliation{Tokyo Institute of Technology, Tokyo}
\affiliation{Tokyo Metropolitan University, Tokyo}
\affiliation{Tokyo University of Agriculture and Technology, Tokyo}
\affiliation{Toyama National College of Maritime Technology, Toyama}
\affiliation{CNP, Virginia Polytechnic Institute and State University, Blacksburg, Virginia 24061}
\affiliation{Wayne State University, Detroit, Michigan 48202}
\affiliation{Yamagata University, Yamagata}
\affiliation{Yonsei University, Seoul}
  \author{I.~Adachi}\affiliation{High Energy Accelerator Research Organization (KEK), Tsukuba} % KEK
  \author{K.~Adamczyk}\affiliation{H. Niewodniczanski Institute of Nuclear Physics, Krakow} % Krakow
  \author{H.~Aihara}\affiliation{Department of Physics, University of Tokyo, Tokyo} % Tokyo
  \author{K.~Arinstein}\affiliation{Budker Institute of Nuclear Physics SB RAS and Novosibirsk State University, Novosibirsk 630090} % BINP
  \author{Y.~Arita}\affiliation{Graduate School of Science, Nagoya University, Nagoya} % Nagoya
  \author{D.~M.~Asner}\affiliation{Pacific Northwest National Laboratory, Richland, Washington 99352} % PNNL
  \author{T.~Aso}\affiliation{Toyama National College of Maritime Technology, Toyama} % Toyama
  \author{V.~Aulchenko}\affiliation{Budker Institute of Nuclear Physics SB RAS and Novosibirsk State University, Novosibirsk 630090} % BINP
  \author{T.~Aushev}\affiliation{Institute for Theoretical and Experimental Physics, Moscow} % ITEP
  \author{T.~Aziz}\affiliation{Tata Institute of Fundamental Research, Mumbai} % Tata
% ...
% BINP
\collaboration{The Belle Collaboration}

\begin{abstract}

We report preliminary results on the analysis of the three-body 
$\UFS\to B\bar{B}\pi$, $\UFS\to [B\bar{B}^*+{\rm c.c.}]\pi$ and 
$\UFS\to B^*\bar{B}^*\pi$ decays including an observation of the 
$\UFS\to Z^{\pm}_b(10610)\pi^\mp\to [B\bar{B}^*+{\rm c.c.}]^{\pm}\pi^\mp$ and 
$\UFS\to Z^{\pm}_b(10650)\pi^\mp\to [B^*\bar{B}^*]^\pm\pi^\mp$ 
decays as intermediate channels. We measure branching fractions of
the three-body decays to be
${\cal{B}}(\UFS\to [B\bar{B}^*+{\rm c.c.}]^{\pm}\pi^{\mp})=
(28.3\pm2.9\pm4.6)\times10^{-3}$ and 
${\cal{B}}(\UFS\to [B^*\bar{B}^*]^{\pm}\pi^{\mp})=
(14.1\pm1.9\pm2.4)\times10^{-3}$
and set 90\% C.L.\, upper limit
${\cal{B}}(\UFS\to [B\bar{B}]^{\pm}\pi^{\mp})<4.0\times10^{-3}$. 
We also report results on the amplitude analysis of the three-body 
$\UFS\to\Un\pp$, $n=1,2,3$ decays and the analysis of the internal 
structure of the three-body $\UFS\to\hm\pp$, $m=1,2$ decays. The 
results are based on a $121.4$~fb$^{-1}$ data sample collected with
the Belle detector at a center-of-mass energy near the $\UFS$. 

\end{abstract}

\maketitle

\section{Introduction}

Two new charged bottomonium-like resonances, $Z_b(10610)$ and $Z_b(10650)$,
have recently been observed by the Belle Collaboration in decays of $\UFS$
to five different final states: $\Upsilon(nS)\pi^+\pi^-$, $n=1,2,3$ and 
$h_b(mP)\pi^+\pi^-$, $m=1,2$~\cite{ypp,hbp}. The analysis of the quark 
composition of the initial and final states allows to assert that these 
hadronic objects are the first examples of states of an exotic nature:
$Z_b$ should be comprised of (at least) four quarks. Several models have
been proposed to describe the internal structure of these states. One  
suggests~\cite{molec} that $Z_b(10610)$ and $Z_b(10650)$ states might be a
loosely bound $B\bar{B}^*$ and $B^*\bar{B}^*$ systems, respectively. The 
proximity of the $Z_b(10610)$ and $Z_b(10650)$ masses to those of the sum of
the $B$ and $B^*$ mesons and the sum of the two $B^*$ mesons, respectively,
supports this hypothesis. In this case, it would be natural to expect that
the $\Zbl$ and $\Zbh$ states decay respectively to $B\bar{B}^*$ and 
$B^*\bar{B}^*$ final states with substantial rates.

Evidence for the three-body $\UFS\to\bbstpi$ decay has been previously 
reported by Belle in Ref.~\cite{drutskoj} with a data sample of
$23.6$~fb$^{-1}$. In this analysis we use $121.4$~fb$^{-1}$ of data 
accumulated by the Belle detector at a center-of-mass (c.m.) energy near
the $\UFS$ to study three-body $\UFS\to [B^{(*)}\bar{B}^{(*)}]^{\pm}\pi^{\mp}$
decays and to search for 
$\UFS\to Z^{\pm}_b\pi^{\mp} \to [B^{(*)}\bar{B}^*]^{\pm}\pi^{\mp}$ decays.

Note that we reconstruct only three-body $B^{(*)}B^{(*)}\pi$ combinations 
with a charged primary pion. For brevity, we adopt the following notations:
the sum of $B^+\bar{B}^0\pi^-$ and $B^-B^0\pi^+$ final states is referred to
as $BB\pi$; the combination of $B^+\bar{B}^{*0}\pi^-$, $B^-B^{*0}\pi^+$, 
$B^0B^{*-}\pi^+$ and $\bar{B}^0B^{*+}\pi^-$ final states is referred to as 
$BB^*\pi$ and the sum of $B^{*+}\bar{B}^{*0}\pi^-$ and $B^{*-}B^{*0}\pi^+$
final states is denoted as $B^*B^*\pi$.

% =============================================================================
% =============================================================================
% =============================================================================

\section{The Belle detector}

The Belle detector~\cite{Belle} is located at the single interaction point of
KEKB~\cite{KEKB}, an asymmetric energy double storage ring collider.
The detector is a large-solid-angle magnetic spectrometer
based on a 1.5~T superconducting solenoid magnet. Charged particle tracking is
provided by a four-layer silicon vertex detector and a 50-layer central
drift chamber (CDC) that surround the interaction point. The charged particle
acceptance covers laboratory polar angles between $\theta=17^{\circ}$ and
$150^{\circ}$, corresponding to about 92\% of the total solid angle in the
c.m.\ frame. 

Charged hadron identification is provided by $dE/dx$ measurements in the CDC,
an array of 1188 aerogel Cherenkov counters (ACC), and a barrel-like array
of 128 time-of-flight scintillation counters (TOF); information from the three
subdetectors is combined to form a single likelihood ratio, which is then used
in kaon and pion selection. Electromagnetic showering particles are
detected in an array of 8736 CsI(Tl) crystals (ECL) that covers the same solid
angle as the charged particle tracking system. 
Electron identification in Belle is based on a combination of $dE/dx$
measurements in the CDC, the response of the ACC, and the position, shape and
total energy deposition ({\it i.e.}, $E/p$) of the shower detected in the ECL.
The magnetic field is returned via an iron yoke that is instrumented to
detect muons and $K^0_L$ mesons. We use a GEANT-based Monte Carlo (MC)
simulation to model the response of the detector and determine its
acceptance~\cite{GEANT}.

% =============================================================================
% =============================================================================
% =============================================================================

\section{Background Suppression}

The dominant background comes from $e^+e^-\to c\bar{c}$ continuum
events where real $D$ mesons produced in $e^+e^-$ annihilation are combined
with random particles to form a $B$ candidate. This type of background is
suppressed using variables that characterize the event topology. Since the
momenta of two $B$ mesons produced from a three-body $\UFS$ decay are low
in the c.m.\ frame, their decay products are essentially uncorrelated and the
event tends to be spherical. In contrast, hadrons from continuum events tend
to exhibit a two-jet structure. We use $\theta_{\rm thr}$, the angle
between the thrust axis of the $B$ candidate and that of the rest of the
event, to discriminate between the two cases. The distribution of 
$|\cos\theta_{\rm thr}|$ is strongly peaked near $|\cos\theta_{\rm thr}|=1.0$
for $c\bar{c}$ events and is nearly flat for $\bbpi$ events. We require
$|\cos\theta_{\rm thr}|<0.80$ for $B\to D^{(*)}\pi$ final states; this
eliminates about 81\% of the continuum background and retains 73\% of the
signal events.

Another significant background comes from events with radiative return to
a lower mass $\Upsilon(4S)$ state with a subsequent $\Upsilon(4S)\to B\bar{B}$ 
decay. Momenta of $B$ mesons produced in this process fall in the same region
as those for $B$ mesons from the three-body $\UFS\to\bbpi$ decays. $B$ mesons
originating from the two-body $\UFS\to B^*B^*$, $BB^*$, $BB$ decays produce
peaks around $P(B)=1.07$~GeV/$c$, 1.18~GeV/$c$ and 1.28~GeV/$c$, respectively.
Momenta of $B$ mesons from three-body $\UFS\to B^{(*)}B^{(*)}\pi$ decays are
less than 0.9~GeV/$c$.

% =============================================================================
% =============================================================================
% =============================================================================

\section{Event Reconstruction}

Charged tracks are selected with a set of track quality requirements based on
the number of CDC hits and on the distances of closest approach to the
interaction point (IP). Tracks originated from $B$ candidate are required
to have momenta transverse to the beam be greater than 0.1~GeV/$c$ to reduce
the low momentum combinatorial background. For charged kaon identification,
we impose a requirement on the particle identification variable, which has 
86\% efficiency and a 7\% fake rate from misidentified pions. Charged tracks
that are positively identified as electrons or protons are excluded. Since
the muon identification efficiency and fake rate vary significantly with the
track momentum, we do not veto muons to avoid additional systematic errors.

Photons from neutral pions are required to produce clusters in the ECL with
an energy deposition of greater than 50~MeV and not be associated with 
charged tracks. The invariant mass of the two-photon combination is required
to be within 12~MeV/$c^2$ of the nominal $\pi^0$ mass. The $K^{*0}$ is 
reconstructed in the $K^{*0}\to K^+\pi^-$ mode, the invariant mass of the 
$K^{*0}$ candidate is required to be within 70~MeV/$c^2$ of the nominal 
$K^{*0}$ mass. The invariant mass of the $J/\psi\to\mu^+\mu^-$ candidates is
required to satisfy $|M(\mu^+\mu^-)-M_{J/\psi}|<30$~MeV/$c^2$, where 
$M_{J/\psi}$
is the nominal mass of the $J/\psi$ meson. Neutral (charged) $D$ mesons 
originating from $B$ decays are reconstructed in the $\bar{D^0}\to K^+\pi^-$
and $\bar{D^0}\to K^+\pi^+\pi^-\pi^-$ ($D^-\to K^+\pi^-\pi^-$) modes. Those 
originating from $D^{*-}$ decays are also reconstructed in the
$\bar{D^0}\to K^+\pi^-\pi^0$ mode. To identify $D^{*-}$ candidates we 
require $|M(\bar{D^0}\pi^-)-M(\bar{D^0})-\Delta M_D|<2$~MeV/$c^2$, where 
$M(\bar{D^0})$ and $M(\bar{D^0}\pi^-)$ are the reconstructed masses of the
$D^0$ candidate and $D^0\pi^-$ system, respectively, and 
$\Delta M_D=M_{D^*}-M_D$.

\begin{figure}[t]
  \includegraphics[width=0.33\textwidth]{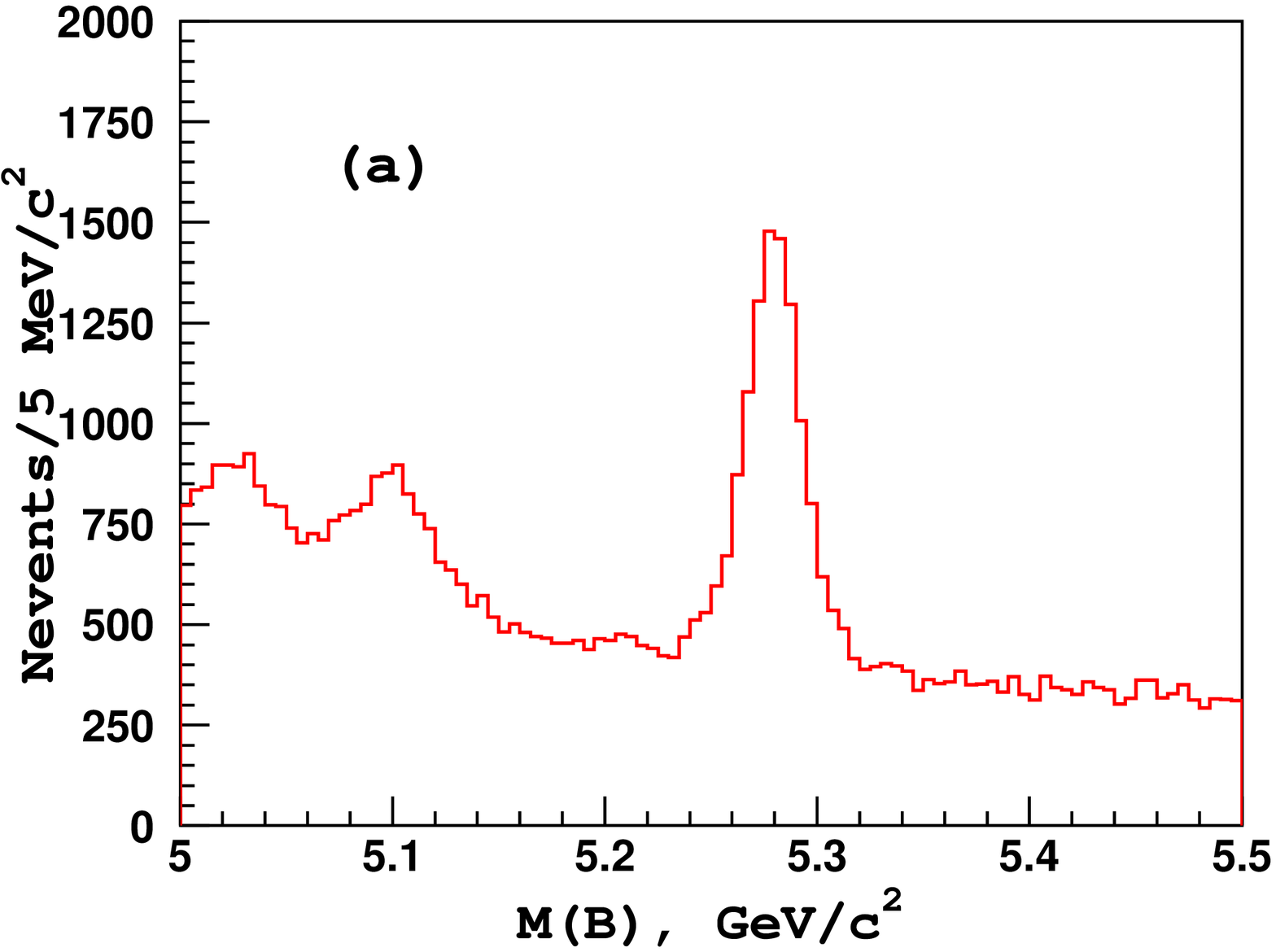} \hspace*{-3mm}
  \includegraphics[width=0.33\textwidth]{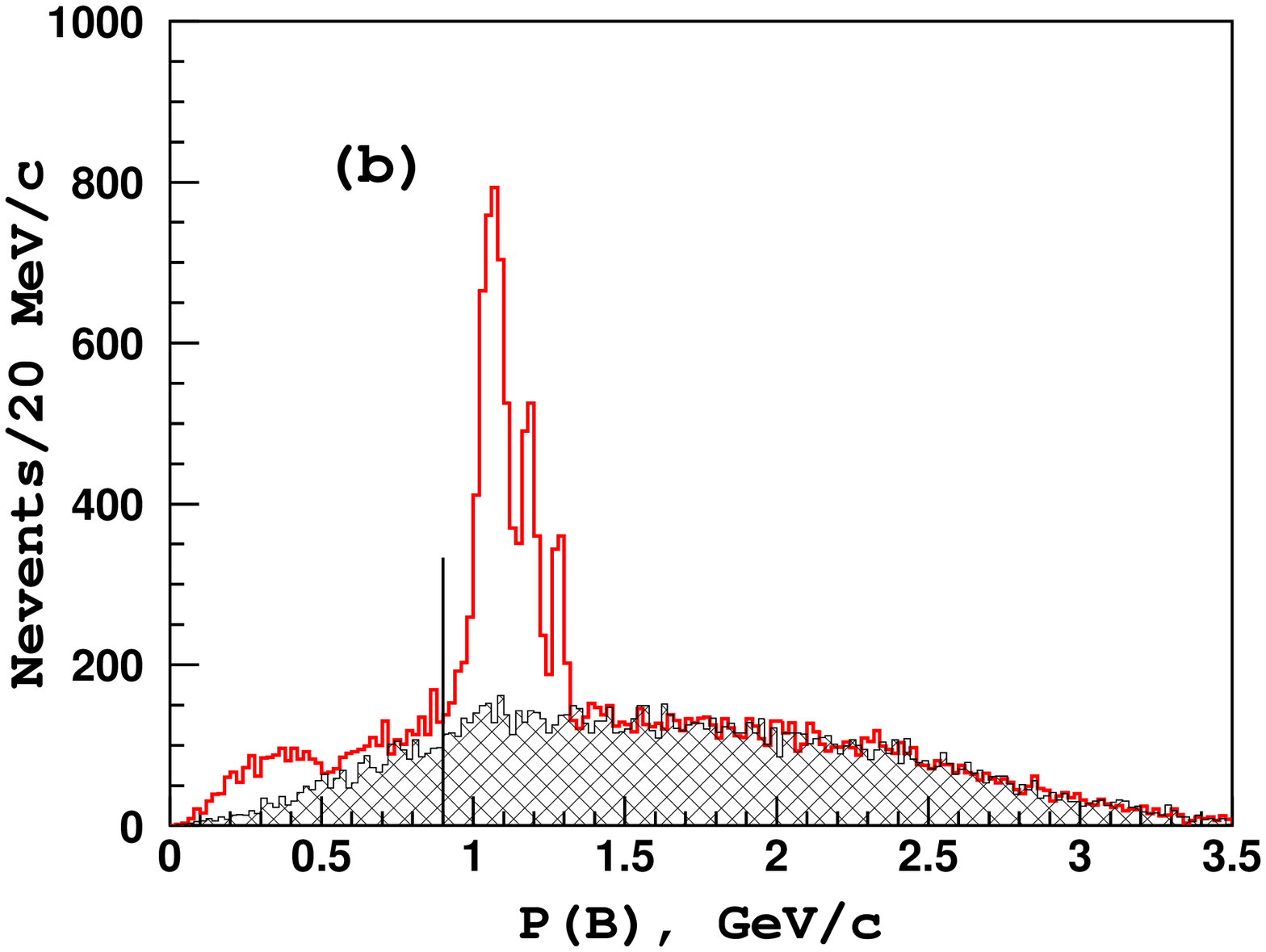} \hspace*{-3mm}
  \includegraphics[width=0.33\textwidth]{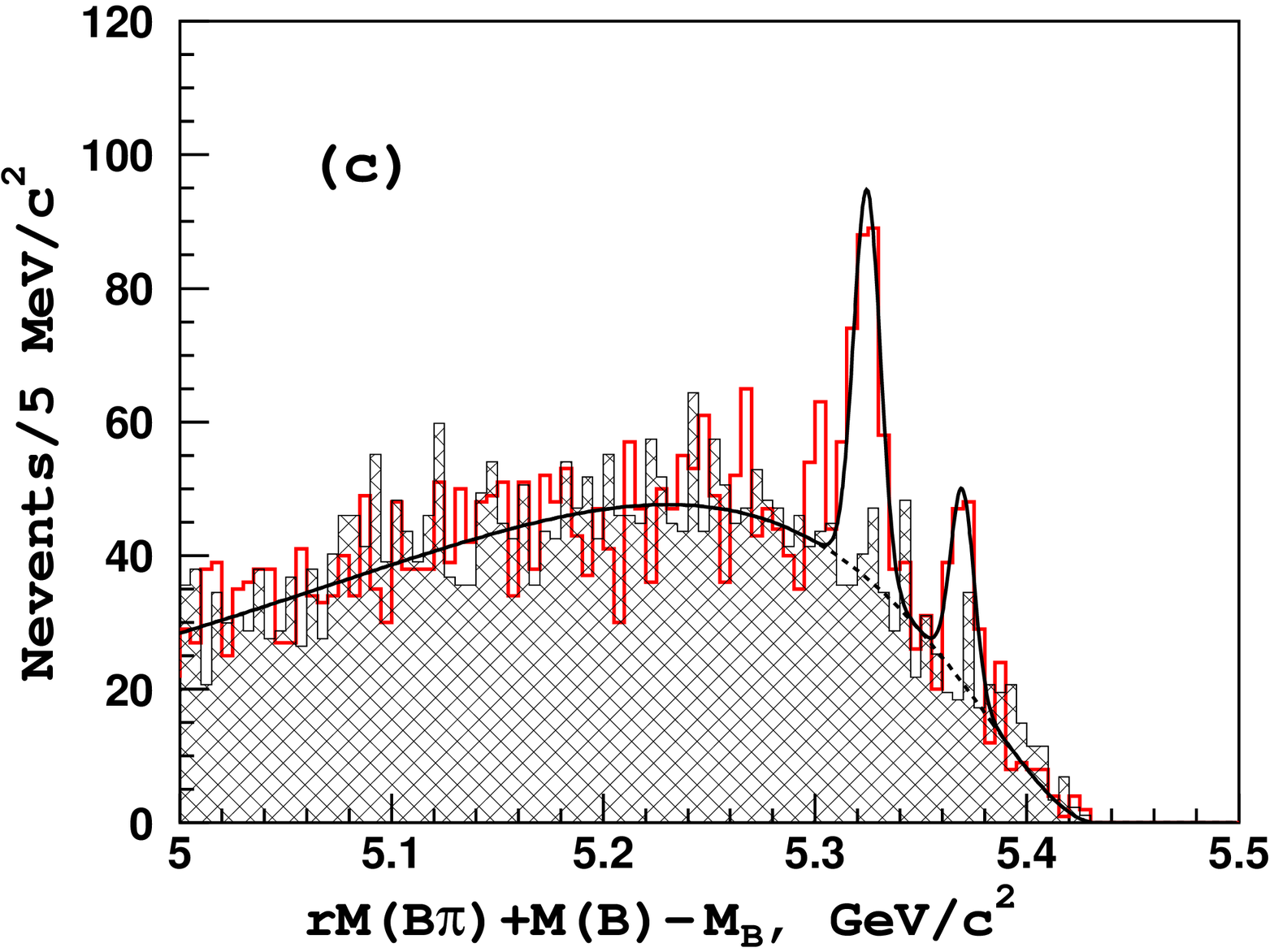}
  \caption{(a) Invariant mass, (b) momentum and (c) $M_r(B\pi)$ distributions
           for selected $B$ candidates in data. Hatched histograms in (b) and
           (c) show distributions for events in $M(B)$ sidebands.}
  \label{fig:select}
\end{figure}

$B$ decays are reconstructed in the following channels:
$B^+\to J/\psi K^+$, $B^+\to \bar{D^0}\pi^+$, $B^0\to J/\psi K^{*0}$,
$B^0\to D^-\pi^+$, $B^0\to D^{*-}\pi^+$. We identify $B$ candidates by their
invariant mass $M(B)$ and momentum $P(B)$. $M(B)$ and $P(B)$ distributions
for $B$ candidates in data are shown in Figs.~\ref{fig:select}(a) 
and~\ref{fig:select}(b). We require $M(B)$ to be within 30 to 40~MeV/$c^2$ 
(depending on the $B$ decay mode) of the nominal $B$ mass. Mass sidebands
are defined as $50$~MeV/$c^2<|M(B)-M_B|<80$~MeV/$c^2$.

Reconstructed $B^+$ or $B^0$ candidates are then combined with a $\pi^-$
candidate and a recoil mass to the $B\pi$ combination, $M_r(B\pi)$, is
calculated as $M_r(B\pi) = \sqrt{E^2_{\rm cms}-P^2_{B\pi}}$, where
$E_{\rm cms}$ is the c.m. energy and $P_{B\pi}$ is the measured three-momentum
of the $B\pi$ combination. Signal $\UFS\to BB^*\pi$ events produce a narrow
peak in the $M_r(B\pi)$ spectrum around the nominal $B^*$ mass, while 
$\UFS\to B^*B^*\pi$ events produce a peak shifted to higher mass by about
45~MeV/$c^2$ due to a missed photon from the $B^*\to B\gamma$ decay. 
It is important to note here that, according to signal MC, $\bbstpi$ events
where the reconstructed $B$ is the one from $B^*$ produce a peak in the
$M_r(B\pi)$ distribution at virtually the same position as $\bbstpi$ events,
where the reconstructed $B$ is the prompt one. To remove a correlation
between $M_r(B\pi)$ and $M(B)$ and to improve the resolution, we use 
$M_r(B\pi)+M(B)-M_B$ instead of $M_r(B\pi)$.
The $M_r(B\pi)+M(B)-M_B$ distribution for experimental data
is shown in Figure~\ref{fig:select}(c), where clear peaks are visible in
the $BB^*\pi$ and $B^*B^*\pi$ signal regions. 

To determine the distribution of background events we combine a 
reconstructed $B$ candidate with pions of the wrong charge. 
The $M_r(B\pi)+M(B)-M_B$ distribution for wrong-sign 
combinations is shown as a hatched histogram in Fig.~\ref{fig:select}(c). 
While wrong-sign $B\pi$ combinations reproduce the shape of the 
combinatorial background very well, the amount of the background is
underestimated by about 18\%. To correct for this effect, we introduce a
scale factor of 1.18 for wrong-sign combinations; the $M_r(B\pi)$ distributions
shown in Fig.~\ref{fig:select}(c) include this correction factor. The 
resolution of the signal peaks in Fig.~\ref{fig:select}(c) is fixed at 
6.1~MeV/$c^2$ as determined from signal MC.

% =============================================================================
% =============================================================================
% =============================================================================

\section{\boldmath Analysis of $\UFS\to [B^{(*)}B^*]^\mp\pi^\pm$}

The fit to the $M_r(B\pi)+M(B)-M_B$ distribution for signal
events shown in Fig.~\ref{fig:select}(c) yields $N_{BB\pi}=1\pm14$, 
$N_{\bbstpi}=184\pm19$ and $N_{\bstbstpi}=82\pm11$ signal events. 
The statistical significance of the observed  $\bbstpi$ and $\bstbstpi$ 
signal is $9.3\sigma$ and $5.7\sigma$, respectively. The statistical 
significance here is calculated as
$\sqrt{-2\ln({\cal L}_0/{\cal L}_{\rm sig})}$, where ${\cal L}_{\rm sig}$ 
and ${\cal L}_0$ denote the likelihood values obtained with the nominal
fit and with the signal yield fixed at zero, respectively.

% =============================================================================

\begin{figure}[!t]
\includegraphics[width=0.49\textwidth]{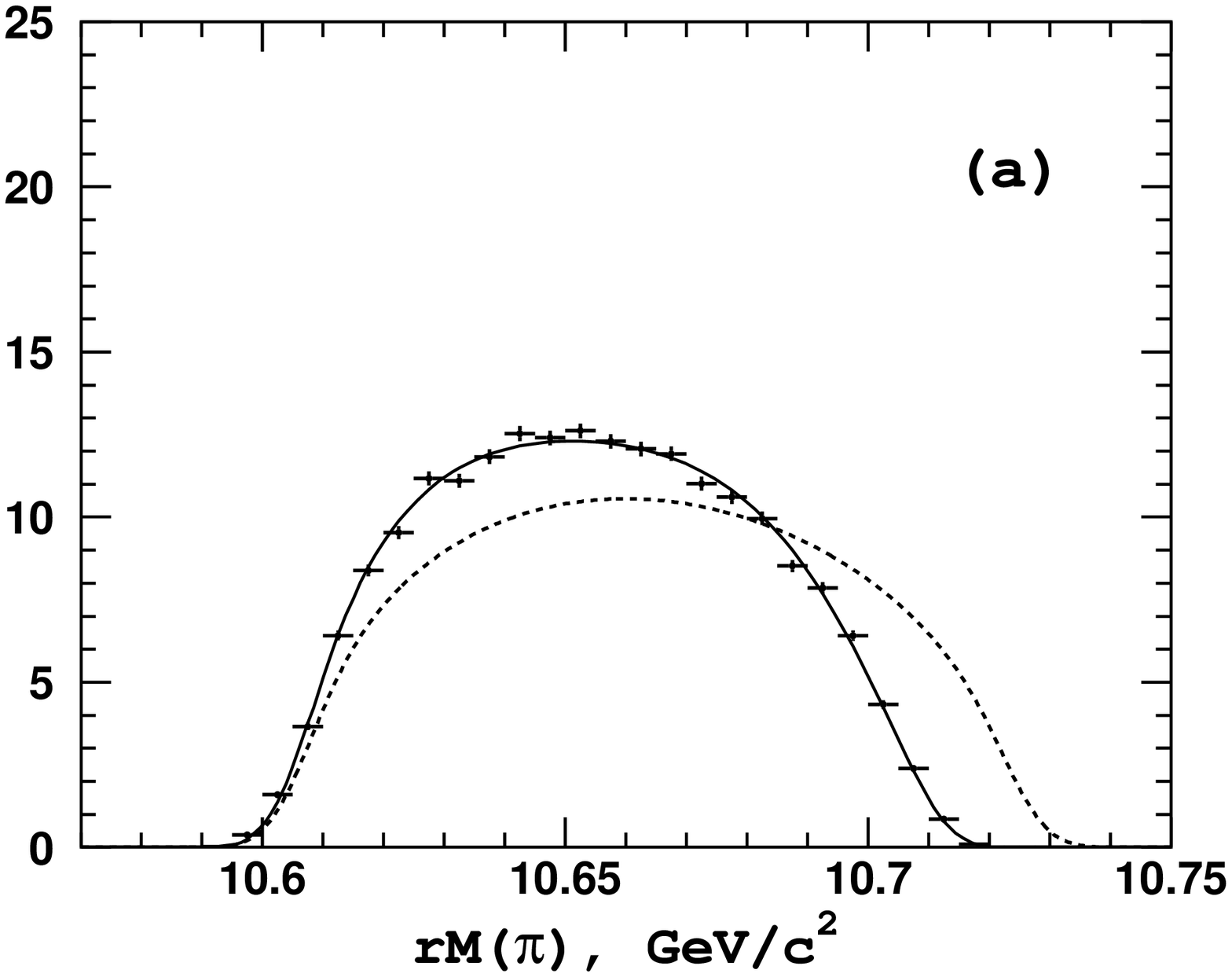} \hfill
\includegraphics[width=0.49\textwidth]{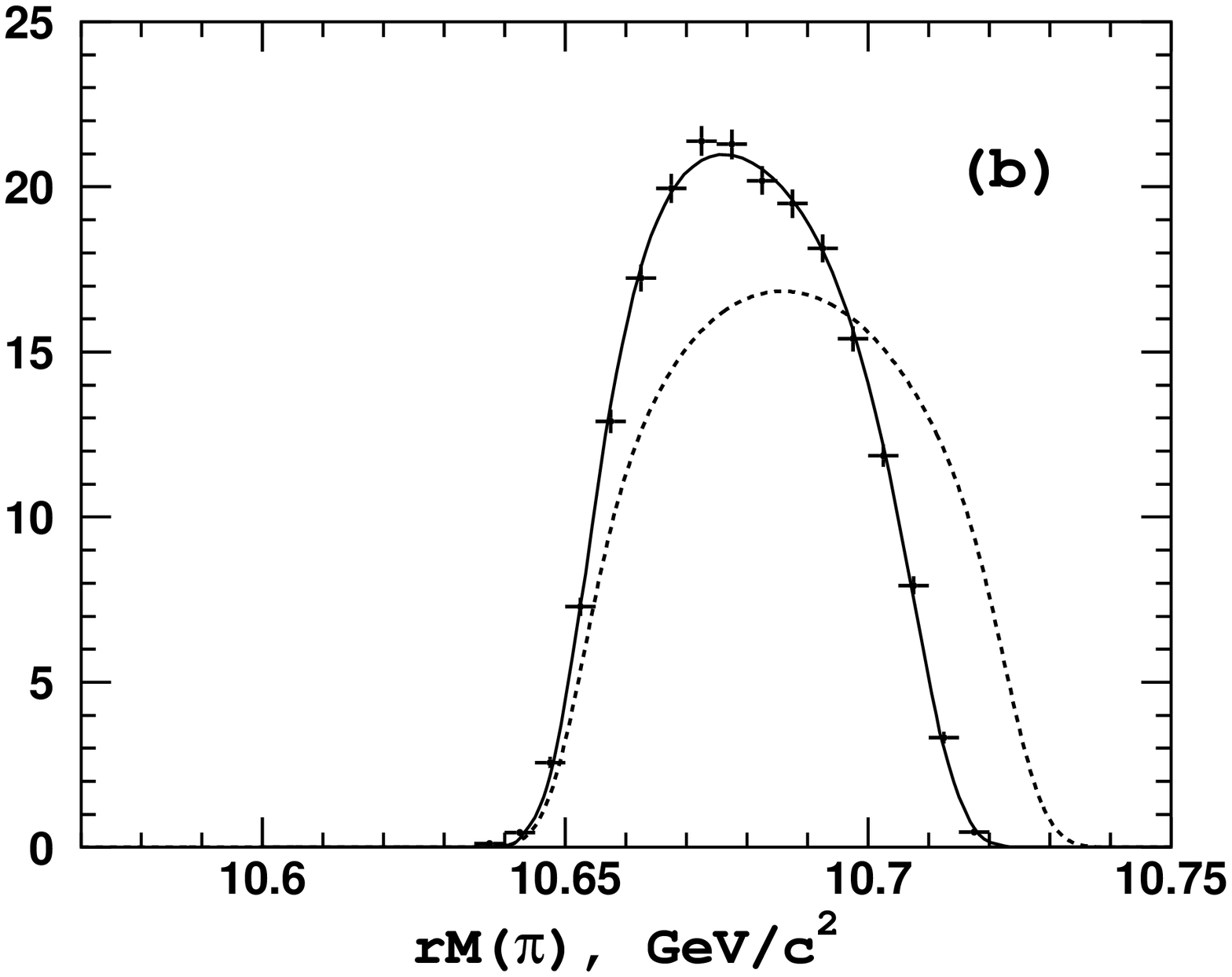}
  \caption{PDFs for reconstruction efficiency for the (a) $\UFS\to\bbstpi$ 
           and (b) $\UFS\to\bstbstpi$ signals. Points with error bars show
           signal MC with the phase-space distribution of signal events, 
           the solid line is the result of the fit, and the dashed line is
           the phase space with uniform efficiency.}
  \label{fig:eff}
\end{figure}

For the subsequent analysis of the internal structures of the three-body
decays, we require $|(M_r(B)+M(B)-M_B)-M_{B^*}|<0.015$~GeV/$c^2$ to select
$\UFS\to\bbstpi$ events and 
$|(M_r(B)+M(B)-M_B)-(M_{B^*}+E_\gamma)|<0.015$~GeV/$c^2$, where 
$E_\gamma=0.049$~GeV~\cite{PDG}, to select $\UFS\to\bstbstpi$ events. For
selected $\bbpi$ candidate events, we
calculate the mass recoiling against the charged pion: 
$M_r(\pi) = \sqrt{E^2_{\rm cms}-P^2_{\pi}}$, where $P_{B\pi}$ is the measured
three-momentum of the charged pion. The $M_r(\pi)$ distributions for signal 
$\UFS\to\bbstpi$ and $\UFS\to\bstbstpi$ MC events generated with the uniform
phase space distribution are shown in Fig.~\ref{fig:eff}. To parameterize
the $M_r(\pi)$ dependence of the reconstruction efficiency $E_{BB^*\pi}(m)$,
we use the following empirical function:
\begin{equation}
E_{\bbpi}(m) = a_0(1+a_1\delta_m+a_2\delta_m^2+a_3\delta_m^3+a_4\delta_m^4)
                 \times{\rm PHSP}_{\bbpi}(m),
\label{eq:bbstp-eff}
\end{equation}
where $m \equiv M_r(\pi)$, $\delta_m=m-m_0$, $m_0=M_{B^{(*)}}+M_{B^*}$ and
$a_i$ are fit parameters, and ${\rm PHSP}(m)$ is the phase space function for
the $\UFS\to\bbpi$ decay. To account for the instrumental resolution, we
smear the efficiency given by Eq.~(\ref{eq:bbstp-eff}) with a Gaussian
function. The resolution $\sigma_m$ of the Gaussian is dominated by the 
c.m.\ energy spread and fixed to be $\sigma_m = 6$~MeV/$c^2$.
The results of the fit are shown in Fig.~\ref{fig:eff}.

\begin{figure}[!t]
\includegraphics[width=0.49\textwidth]{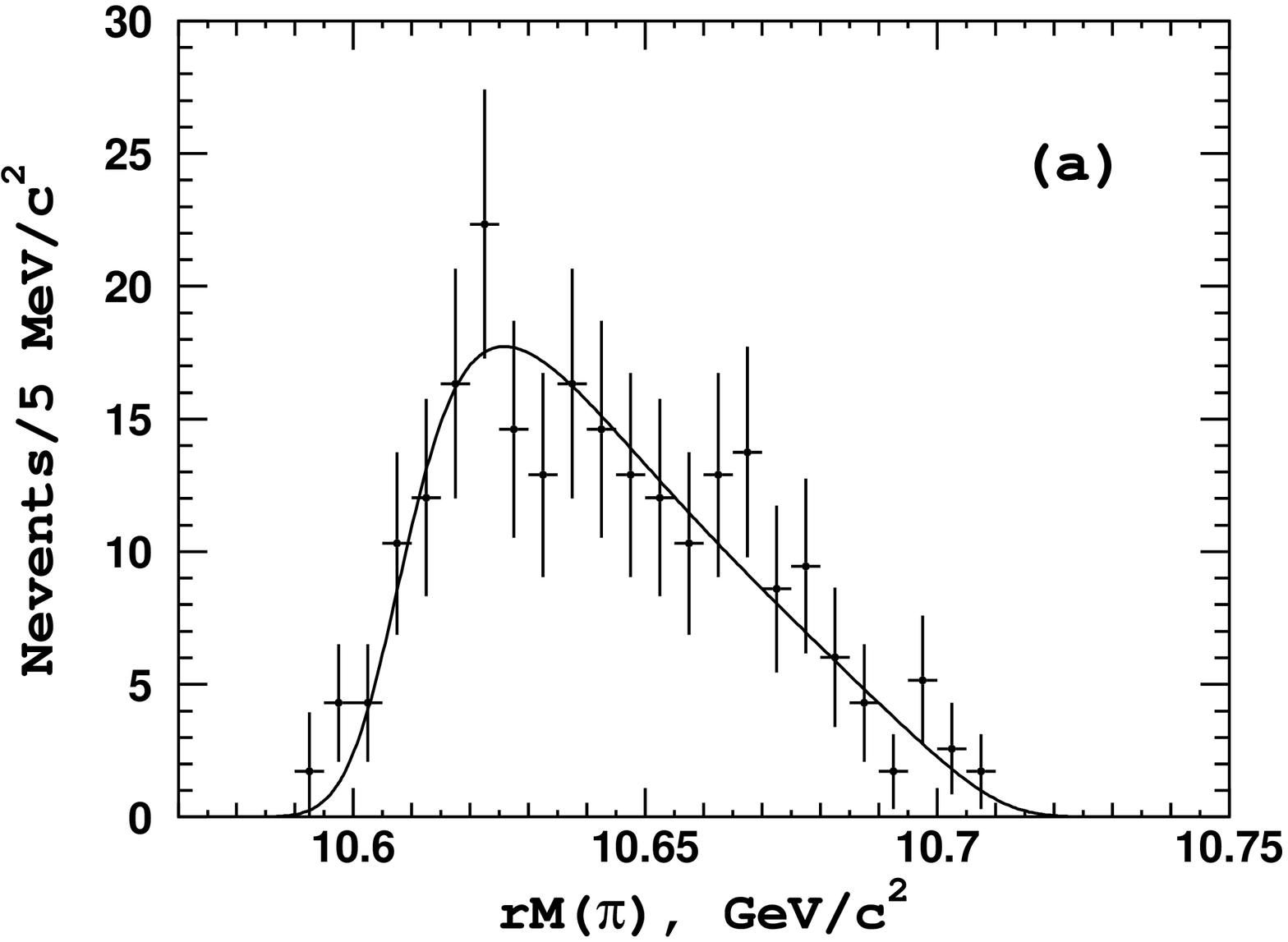} \hfill
\includegraphics[width=0.49\textwidth]{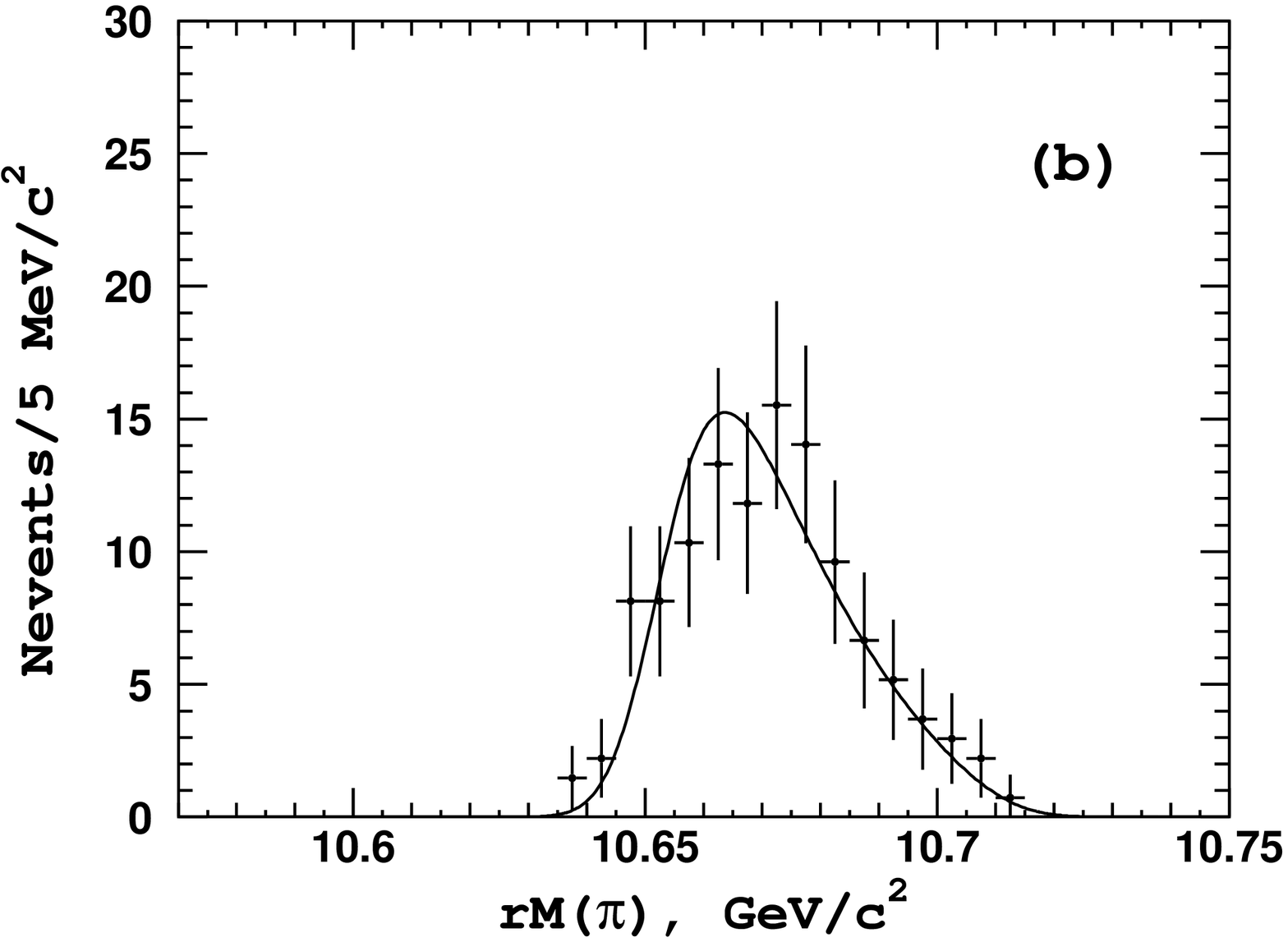}
  \caption{$M_r(\pi)$ distribution for wrong-sign $B\pi$ combinations for the
           (a) $\bbstpi$ and (b) $\bstbstpi$ candidate events.
           Points with error bars are data, the solid line is the result of 
           the fit with a function of Eq.(\ref{eq:bkg}).}
  \label{fig:bkg}
\end{figure}

% =============================================================================

The $M_r(\pi)$ distributions for wrong-sign $B\pi$ combinations for events in
the $BB^*\pi$ and $B^*B^*\pi$ signal regions are shown in Fig.~\ref{fig:bkg}.
We use the following empirical function to parameterize the distribution of
background events
\begin{equation}
B_{B^{(*)}B^*\pi}(m) = b_0e^{-\alpha\delta_m}\times E_{B^{(*)}B^*\pi}(m),
\label{eq:bkg}
\end{equation}
where $b_0$ and $\alpha$ are fit parameters. As in the case of the fit 
to the efficiency PDF, the background PDF is convolved with a resolution 
function. Results of fits to sideband events are shown in Fig.~\ref{fig:bkg}.

\begin{figure}[!t]
\includegraphics[width=0.49\textwidth]{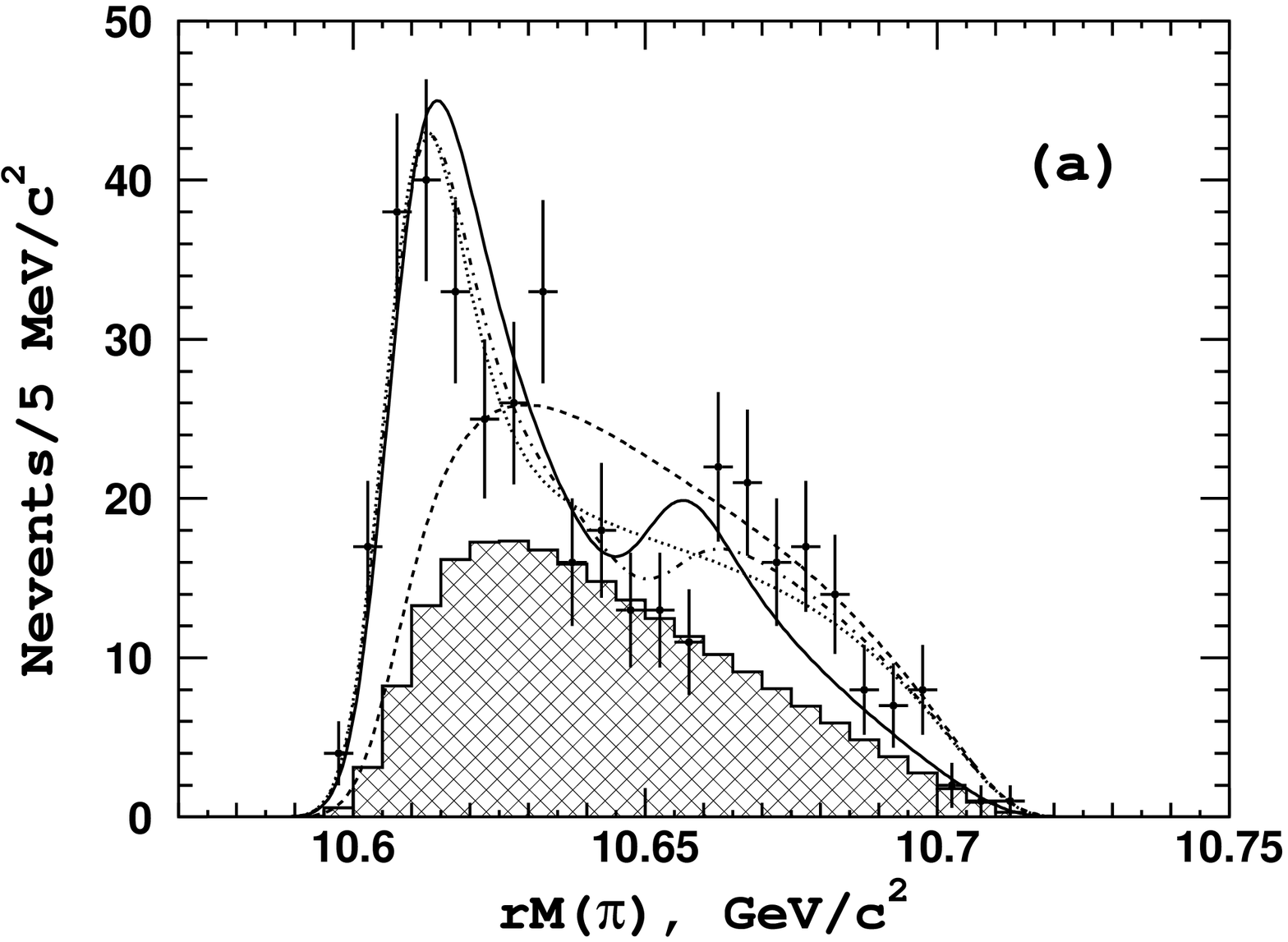} \hfill
\includegraphics[width=0.49\textwidth]{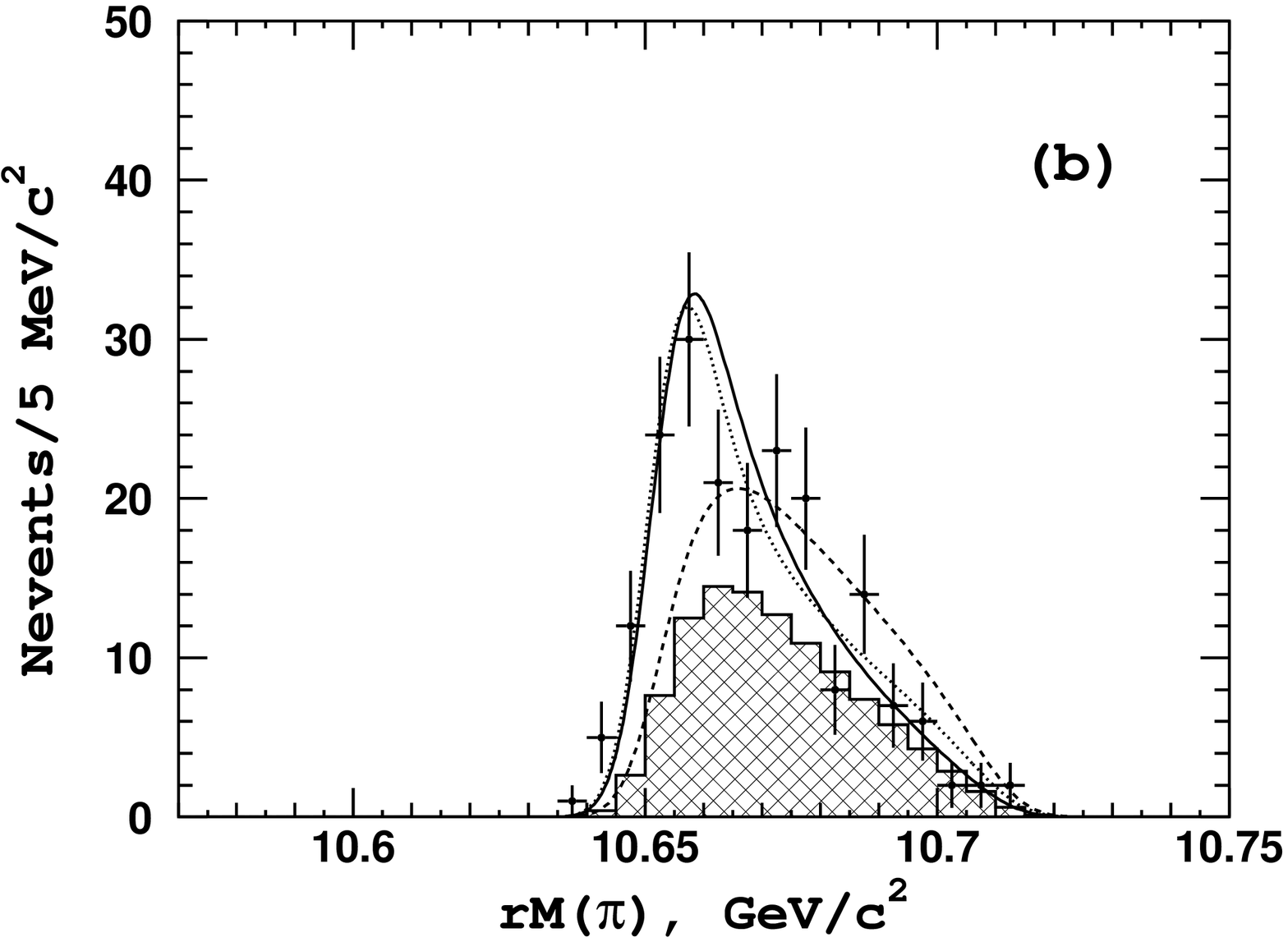}
  \caption{$M_r(\pi)$ distribution for right-sign $B\pi$ combinations for 
           (a) $\UFS\to\bbstpi$ and (b) $\UFS\to\bstbstpi$ candidate events.
           Points with error bars are data, the solid line is the result of the
           fit with the nominal model (see text), the dashed line - fit to pure 
           non-resonant amplitude, the dotted line - fit to a single $Z_b$
           state plus a non-resonant amplitude, and the dash-dotted - two 
           $Z_b$ states and a non-resonant amplitude. The hatched histogram
           represents background component normalized to the estimated number
           of background events.}
  \label{fig:sig}
\end{figure}

% =============================================================================

The $M_r(\pi)$ distributions for right-sign $B\pi$ combinations in the 
$BB^*\pi$ and $B^*B^*\pi$ signal regions are shown in Fig.~\ref{fig:sig}. 
Excesses of signal events over the expected background levels at lower mass
edges of the $M_r(\pi)$ spectra are clearly visible for both final states.
The distribution of signal $\UFS\to\bbstpi$ events is parameterized with
the following model
\begin{eqnarray}
S_{BB^*\pi}(m) = (A_{Z_b(10610)} + A_{NR})\times E_{BB^*\pi}(m),
\label{eq:bbstp-sign}
\end{eqnarray}
where $A_{NR}$ is the non-resonant amplitude parameterized as a complex
constant and the $Z_b(10610)$ amplitude is a Breit-Wigner function. As a 
variation of this nominal model, we also add a second Breit-Wigner amplitude
to account for possible $Z_b(10650)\to\bbstpi$ decay. We also fit the data
with only the $Z_b(10610)$ channel included in the decay amplitude. 
The results of these fits are shown in Fig.~\ref{fig:sig}(a). 
Two models give about equally good description of the data: 
nominal model and a model with additional non-resonant amplitude.
However, we select the former one as our nominal model since adding a 
non-resonant amplitude does not improve the fit quality that much. The worst
fit to the data is provided by a model with just a non-resonant amplitude.
From this analysis, we estimate that the significance of the $\Zbl\to BB^*$
signal exceeds the $8\sigma$ level.

As the nominal model for the $\UFS\to\bstbstpi$ decay, we use the following
parameterization:
\begin{eqnarray}
S_{B^*B^*\pi}(m) = (A_{Z_b(10650)} + A_{NR}){\rm E}_{B^*B^*\pi}(m).
\label{eq:bstbstp-sign}
\end{eqnarray}
We also fit the data without a non-resonant component and with a non-resonant
amplitude alone. Results of the fits are shown in Fig.~\ref{fig:sig}(b);
numerical values are given in Table~\ref{tab:results}. 

The best description of the $\bstbstpi$ data is achieved in a model with only
the $\Zbh$ amplitude included. The addition of a non-resonant amplitude does
not provide any significant improvement of the fit quality. The fit with a 
non-resonant amplitude alone gives a much worse likelihood value. From this
analysis, we determine the significance of the $\Zbh\to B^*B^*$ signal to be
$6.8\sigma$.

\begin{table}[!t]
\centering
\caption{Summary of fit results to the $M_r(\pi)$ distribution for
         three-body $\UFS\to\bbstpi$ and $\UFS\to\bstbstpi$ decays.}
\medskip
\label{tab:results}
  \begin{tabular}{l|cccccc}  \hline \hline
 Mode ~~~& Parameter & 
~~~Nominal model~~~ & ~~~Model-1~~~ & ~~~Model-2~~~ & ~~~Model-3~~~ & ~~~Model-4~~~
\\
\hline 
$\bbstpi$
& $f_{\Zbl}$      & $1.08\pm0.12$ &  $-$  & $0.86\pm0.15$ & $1.0$ & $0.73\pm0.17$  \\
& $f_{\Zbh}$      & $0.25\pm0.10$ &  $-$  &      $-$      &  $-$  & $0.087\pm0.061$\\
& $\phi_{\Zbh}$   & $-0.93\pm0.34$&  $-$  &      $-$      &  $-$  & $0.32\pm0.23$  \\
& $f_NR$          & $-$           & $1.0$ & $1.37\pm0.28$ &  $-$  & $1.17\pm0.27$  \\
& $\phi_{NR}$     & $-$           &  $-$  & $0.18\pm0.21$ &  $-$  & $2.75\pm1.03$  \\
& $-\log{\cal{L}}$& $142$         & $226$ &    $129$     & $162$ & $126$           \\
\hline
$\bstbstpi$
& $f_{\Zbh}$      &    $1.0$      &      $-$      & $0.83\pm0.14$ &   \\
& $f_NR$          &     $-$       &     $1.0$     & $0.78\pm0.43$ &   \\
& $\phi_{NR}$     &     $-$       &      $-$      & $0.53\pm2.4$  &   \\
& $-\log{\cal{L}}$&    $86.0$     &    $133.6$    &    $83.6$     &   \\
\hline \hline
  \end{tabular}
\end{table}

In all fits discussed above, the masses and widths of the $Z_b$ states were
fixed at the values obtained from the analysis of the $\Upsilon(nS)\pp$ and 
$h_b(mP)\pp-$ final states: 
$M[\Zbl] = 10607.2\pm2.0$~MeV/$c^2$, $\Gamma[\Zbl]=18.4\pm2.4$~MeV and
$M[\Zbh] = 10652.2\pm1.5$~MeV/$c^2$, $\Gamma[\Zbh]=11.5\pm2.2$~MeV.
If allowed to float, the fit returns $10597\pm9$~MeV for $\Zbl$ mass in the
fit to $\bbstpi$ events and $10649\pm12$~MeV for $\Zbh$ mass in the fit to
$\bstbstpi$ events. Large errors here reflect a strong negative correlation
between resonance mass and its amplitude.

% =============================================================================
% =============================================================================
% =============================================================================

\section{\boldmath Analysis of $\Uf\to\Un\pp$}

In addition to new results on the analysis of the three-body $\UFS\to\bbpi$
decays described in previous sections, we extend our analysis of the
three-body $\UFS\to\Un\pp$ and $\UFS\to\hm\pp$ decays reported earlier 
in Ref.~\cite{ypp} to measure not only the parameters of the newly 
observed $Z_b$ states but also the fractions of individual components 
contributing to the three-body signals.

To select $\Uf\to\Un\pp$ ($n=1,2,3$) candidate events, we require the 
presence of a pair of muon candidates with an invariant mass in the range of 
$8.0$~GeV/$c^2<M(\mu^+\mu^-)<11.0$~GeV/$c^2$ and two pion candidates of
opposite charge in the event. All tracks are required to originate from the
vicinity of the interaction point. We also require that none of the four
tracks be consistent with being an electron. More details on the analysis
flow can be found in Ref.~\cite{ypp} and references therein. 

\begin{figure}[!t]
  \centering
\hspace*{-1mm}
  \includegraphics[width=0.32\textwidth]{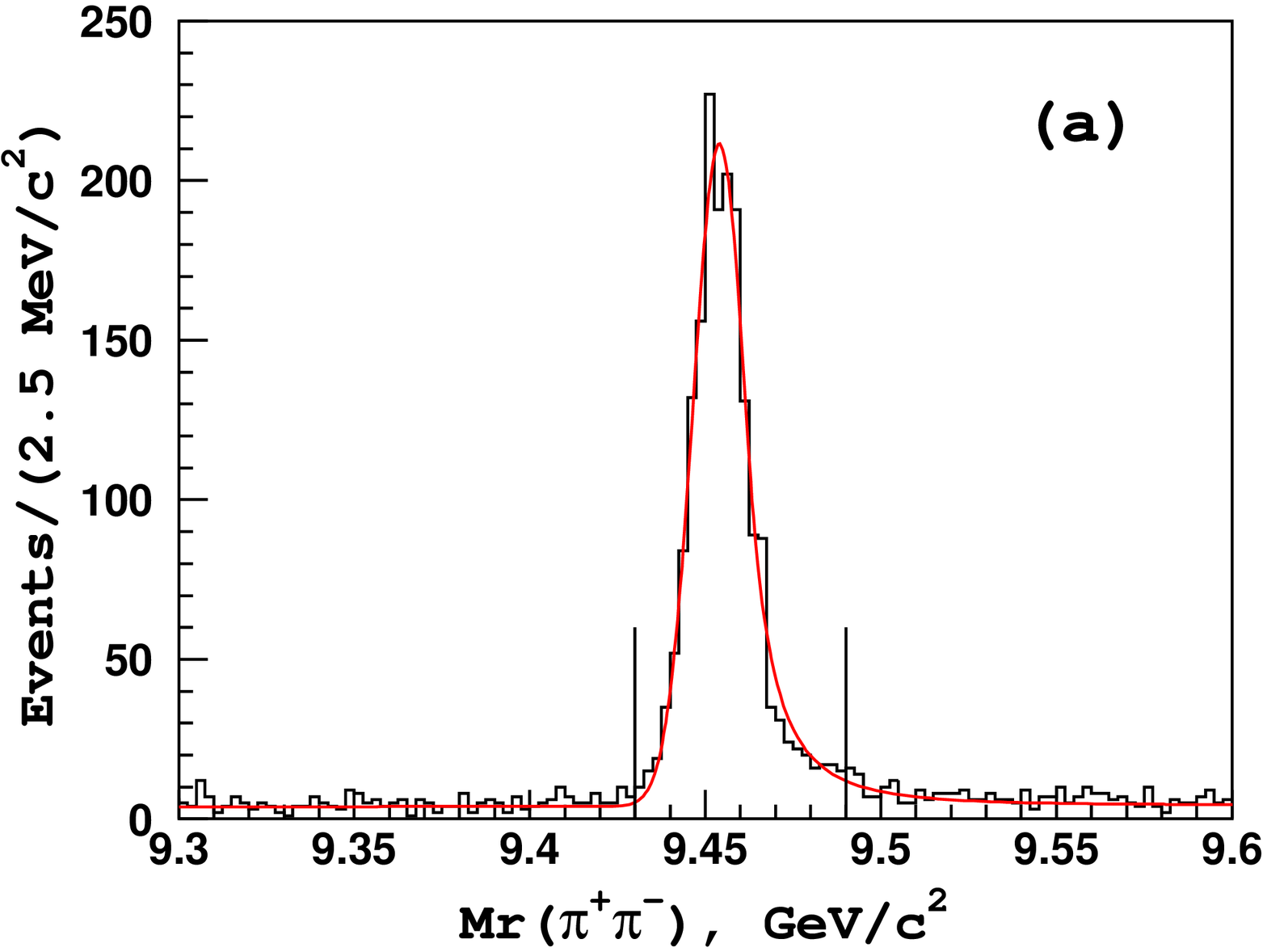} \hfill
  \includegraphics[width=0.32\textwidth]{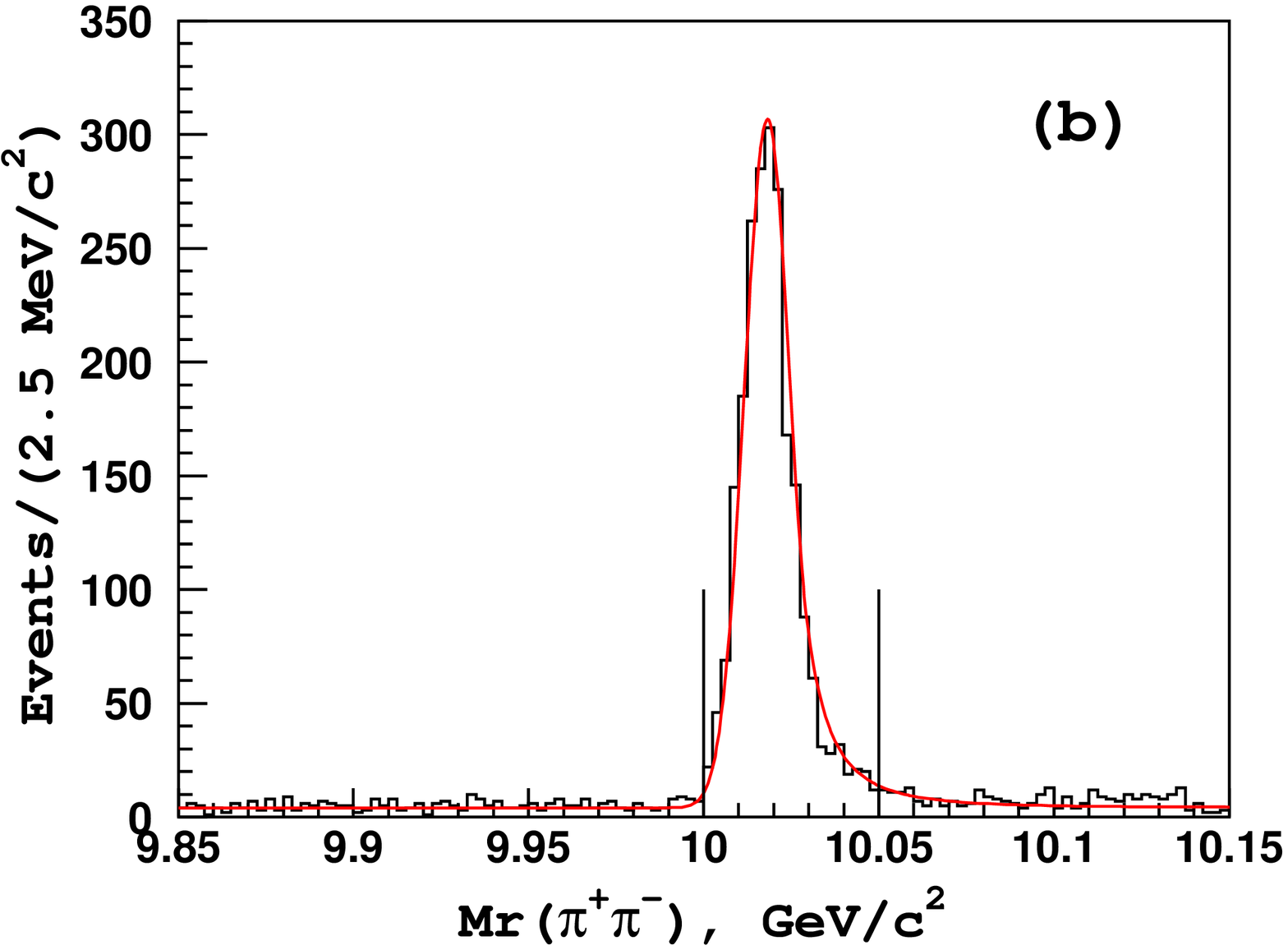} \hfill
  \includegraphics[width=0.32\textwidth]{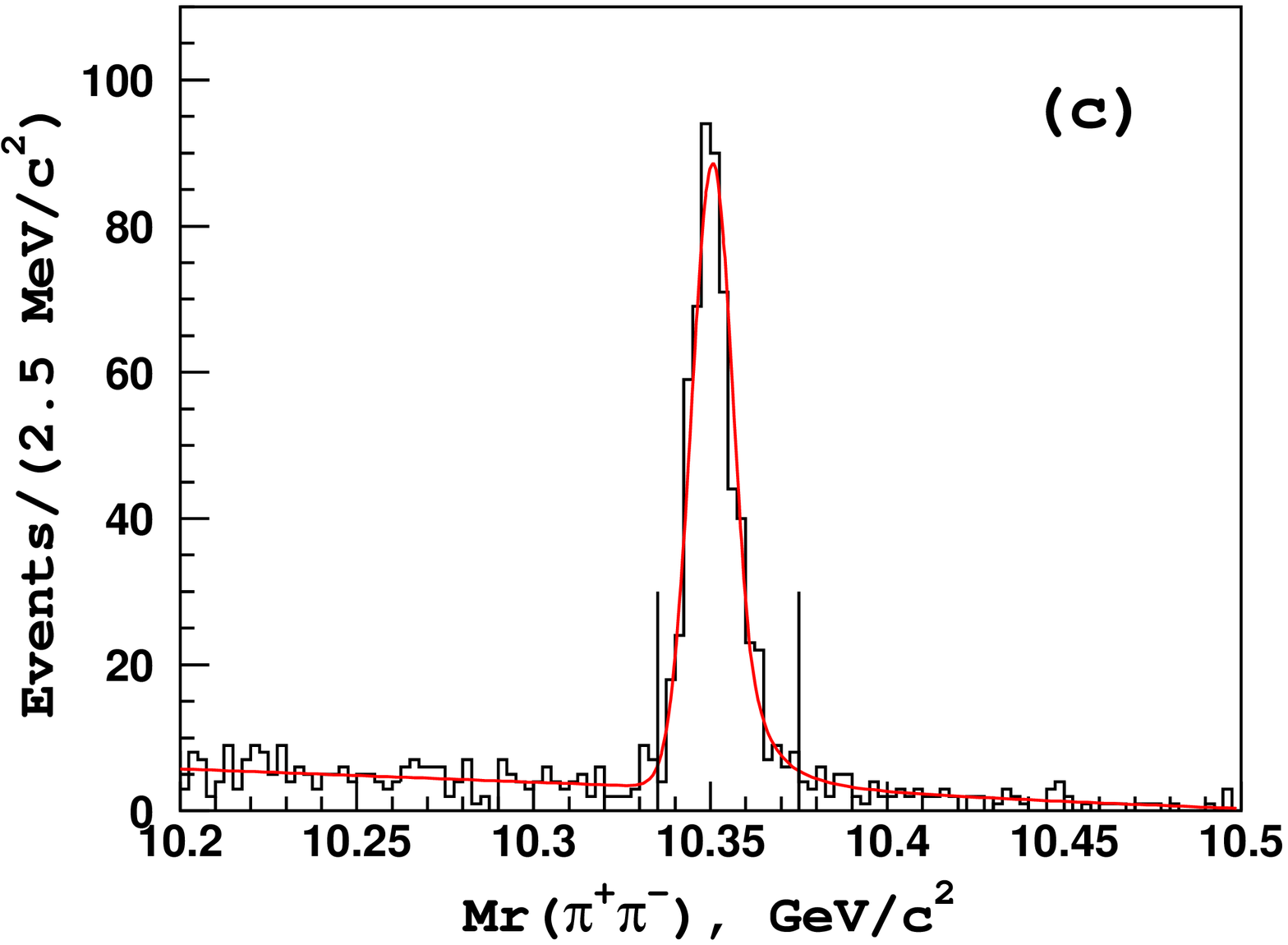}
  \caption{Distribution of recoil mass associated with the $\pp$ 
           combination for
           $\Upsilon(5S)\to\Upsilon(nS)\pi^+\pi^-$ candidate events in the
           (a) $\Upsilon(1S)$; (b) $\Upsilon(2S)$; (c) $\Upsilon(3S)$ mass
           region. Vertical lines define the corresponding signal region.}
\label{fig:ynspp-s-mm}
\end{figure}

Candidate $\Uf\to\Un\pp$ events are identified via the measured invariant 
mass of the $\uu$ combination and the recoil mass, $M_r(\pp)$, associated
with the $\pp$ system calculated as
$M_r(\pp)\equiv\sqrt{(E_{\rm c.m.}-E_{\pp}^*)^2-p_{\pp}^{*2}}$, where 
$E_{\pp}$ and 
$p_{\pp}^*$ are the energy and momentum of the $\pp$ system measured in
the c.m.\ frame. Events originating from $\Uf$ decays are selected with the
requirement of $|M_r(\pp)-M(\uu)|<0.2$~GeV/$c^2$. The $M_r(\pp)$ distributions
shown in Fig.~\ref{fig:ynspp-s-mm} are fit to the sum of a Crystal Ball 
function for the $\Un$ signal and a linear function for the combinatorial
background component. 
Results of the fits are shown in Fig.~\ref{fig:ynspp-s-mm}.

For the subsequent analysis, we select events around a respective $\Un$ mass 
peak
as shown in Fig.~\ref{fig:ynspp-s-mm}. After all the selections are applied,
we come up with $1819$, $2219$ and $588$ events for the $\Uo\pp$, $\Ut\pp$
and $\Uh\pp$ final state, respectively. The fractions of signal events in the 
selected samples are determined from the fit to the corresponding $M_r(\pp)$
spectrum.

The amplitude analysis of three-body $\Uf\to\Un\pp$ ($n=1,2,3$) decays is
performed by means of an unbinned maximum likelihood fit. The distribution
of background events is determined using events in the $\Un$ mass sidebands.
The variation of reconstruction efficiency over the phase space is determined
using MC simulated signal events generated with a uniform distribution.

We use the following parameterization of the $\Uf\to\Un\pp$ three-body decay
amplitude:
\[
M(s_1,s_2) = A_1(s_1,s_2) + A_2(s_1,s_2) + A_{f_0} + A_{f_2} + A_{NR},
\]
where $s_1 = m^2(Y(nS)\pi^+)$, $s_2 = m^2(Y(nS)\pi^-)$. The amplitudes $A_1$
and $A_2$ are $S$-wave Breit-Wigner functions to account for the observed 
$Z_b(10610)$ and $Z_b(10650)$ peaks, respectively. To account for the
possibility for the $\Uf$ to decay to both $Z^+\pi^-$ and $Z^-\pi^+$ channels,
the amplitudes $A_1$ and $A_2$ are symmetrized with respect to $\pi^+$ and
$\pi^-$ interchange. Taking into account isospin symmetry, the resulting
amplitude is written as
\[
A_k = a_k e^{i\delta_k} (BW(s_1,m_k,\Gamma_k) + BW(s_2,m_k,\Gamma_k)),
\]
where the masses $m_k$ and widths $\Gamma_k$ ($k = 1,2$) are free parameters
of the fit. Due to the very limited phase space available  in $\Uf\to\Un\pp$
decays, $\Uf\to Z^+_{b}\pi^-$ and $\Uf\to Z^-_{b}\pi^+$ amplitudes overlap
significantly. We also include amplitudes $A_{f_0}$ and $A_{f_2}$ to account
for possible contributions from $f_0(980)$ scalar and $f_2(1270)$ tensor
states. We use a Breit-Wigner function to parameterize the $f_2(1270)$ and
a Flatte function for the $f_0(980)$. The mass and width of the $f_2(1270)$ 
state are fixed at their world average values~\cite{PDG}; the mass and coupling
constants of the $f_0(980)$ state are fixed at the values defined from the
analysis of $B^+\to K^+\pi^+\pi^-$: $M(f_0(980))=950$~MeV/$c^2$, 
$g_{\pi\pi}=0.23$, $g_{KK}=0.73$~\cite{kpp}.

\begin{figure}[!t]
  \centering
  \includegraphics[width=0.32\textwidth]{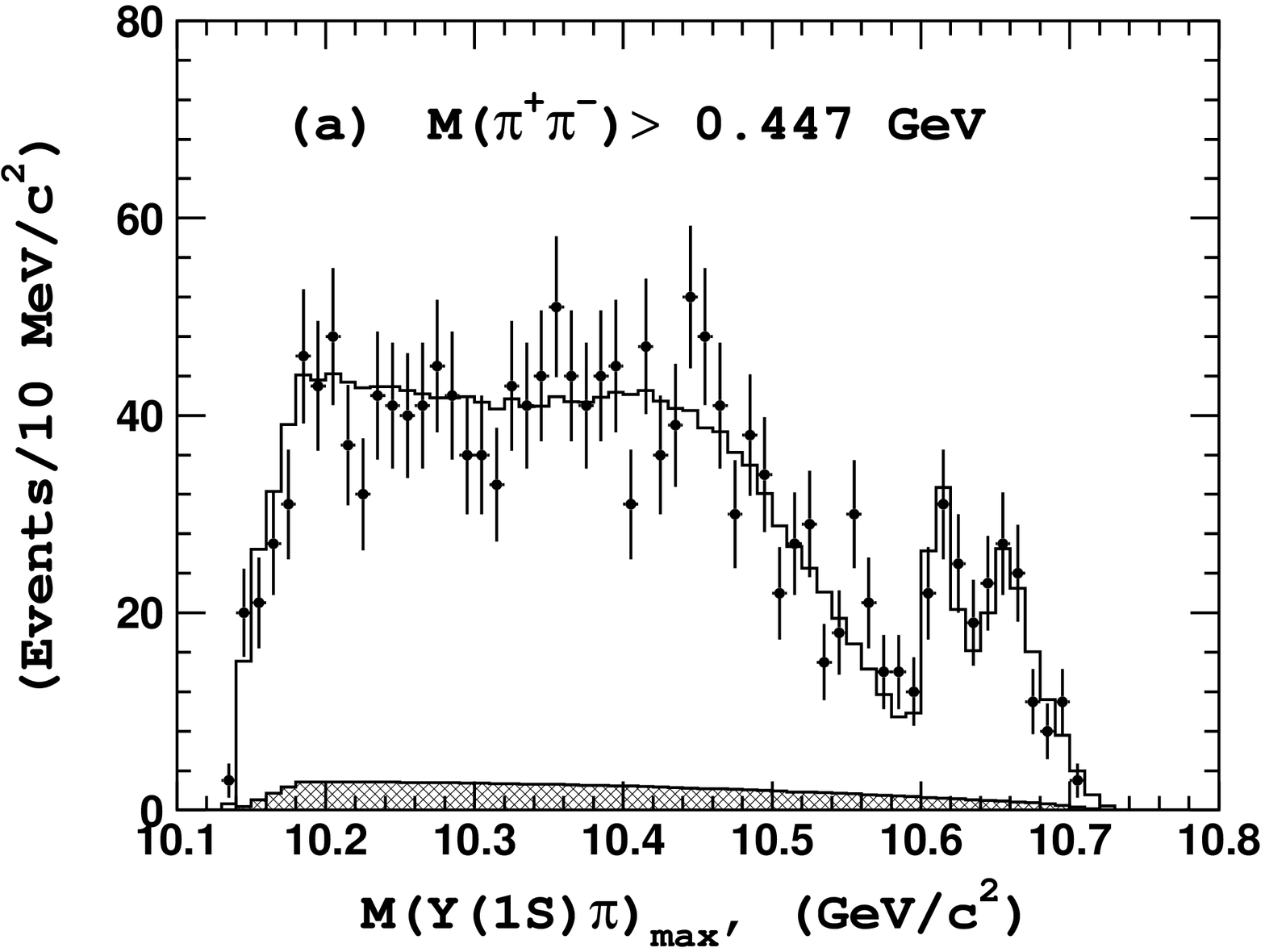} \hfill
  \includegraphics[width=0.32\textwidth]{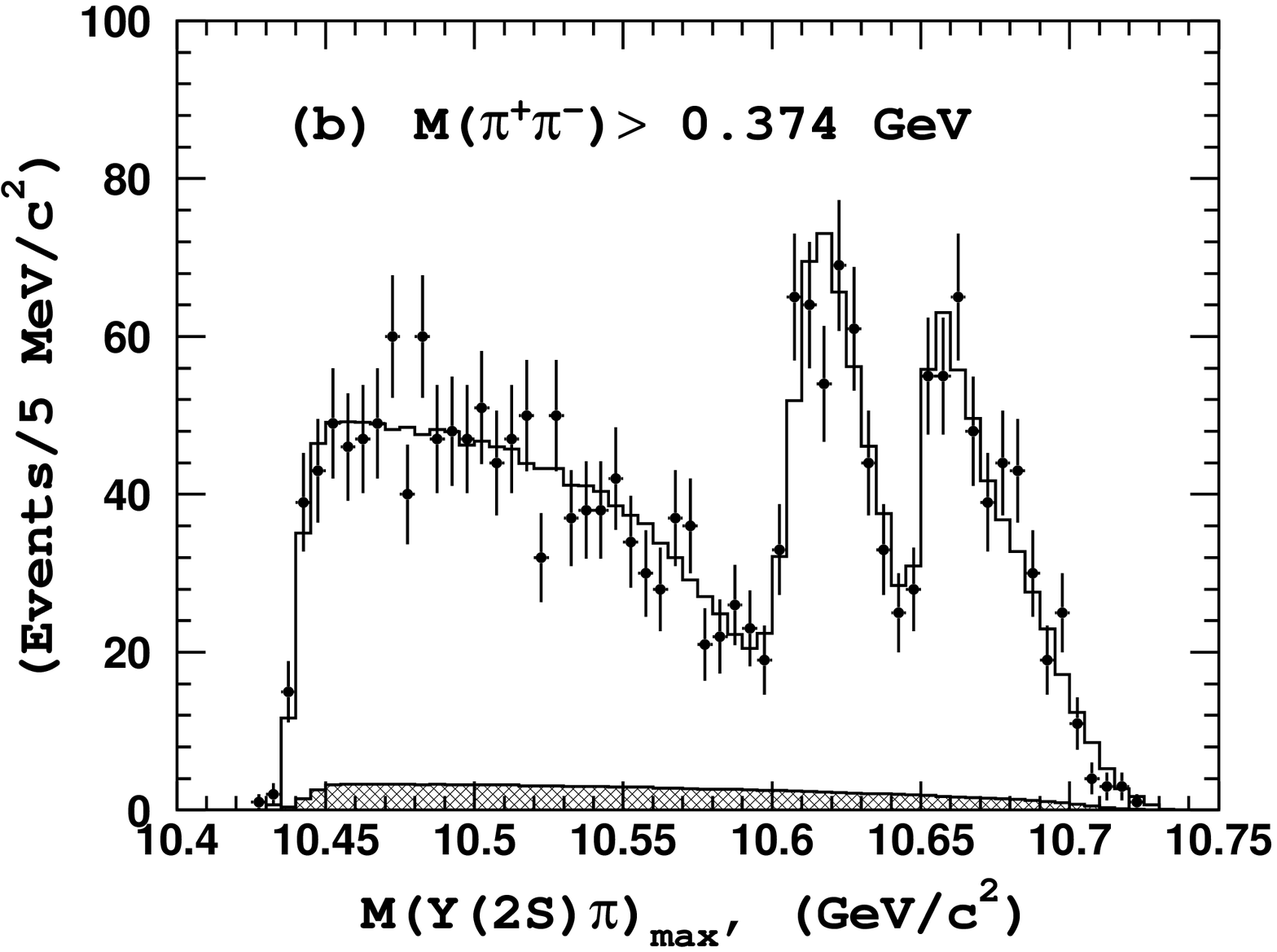} \hfill
  \includegraphics[width=0.32\textwidth]{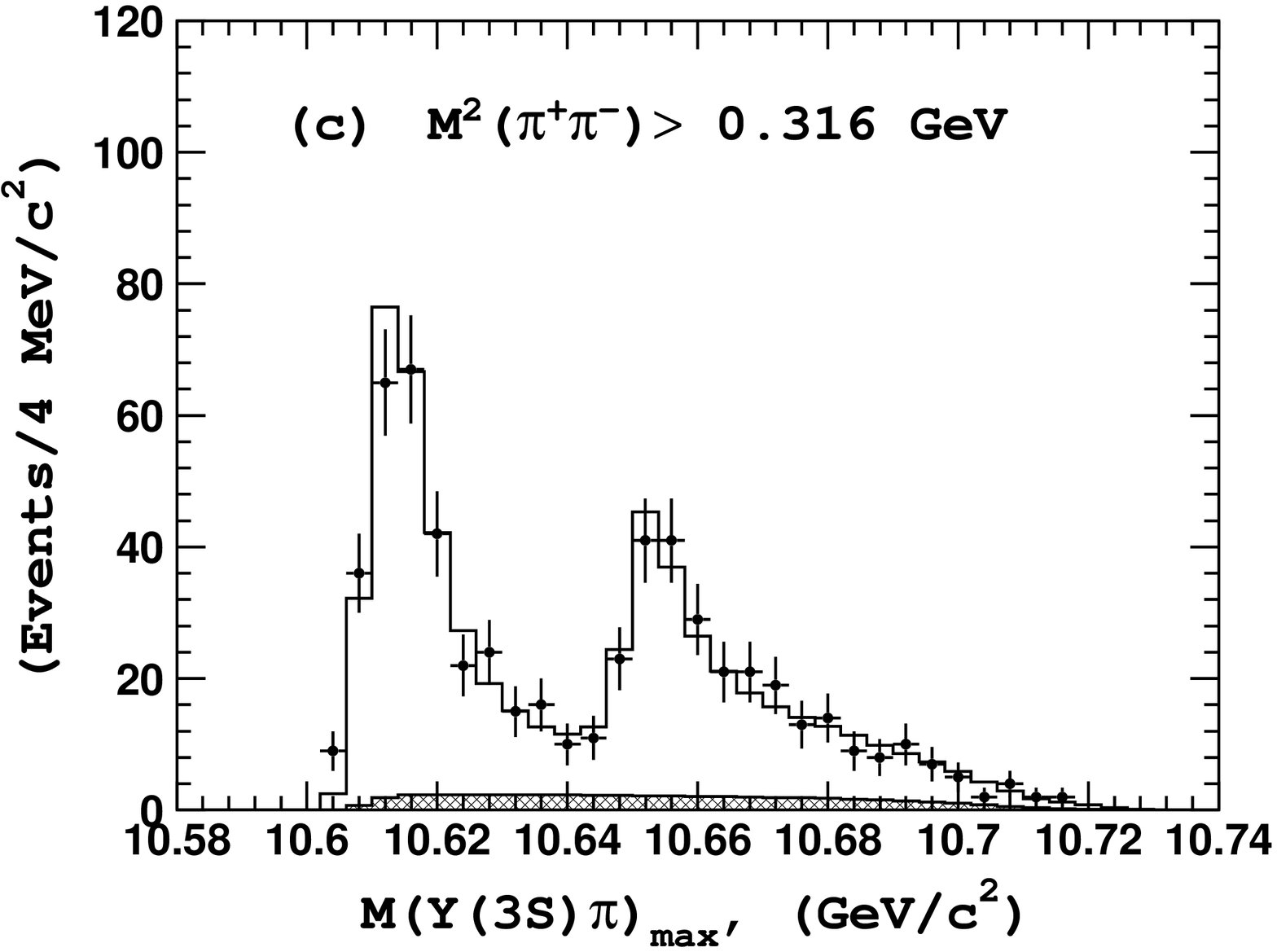} \\
  \includegraphics[width=0.32\textwidth]{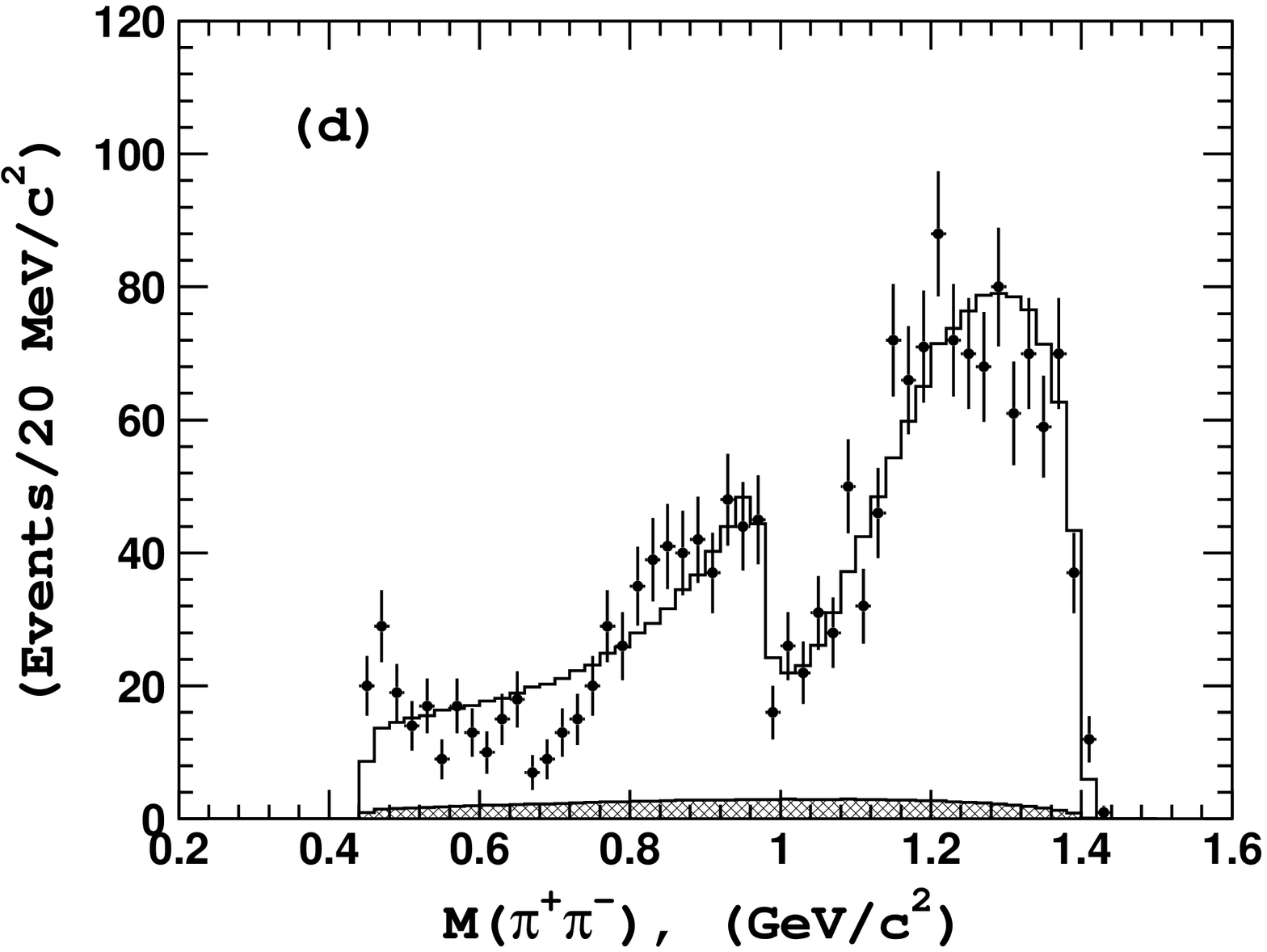} \hfill
  \includegraphics[width=0.32\textwidth]{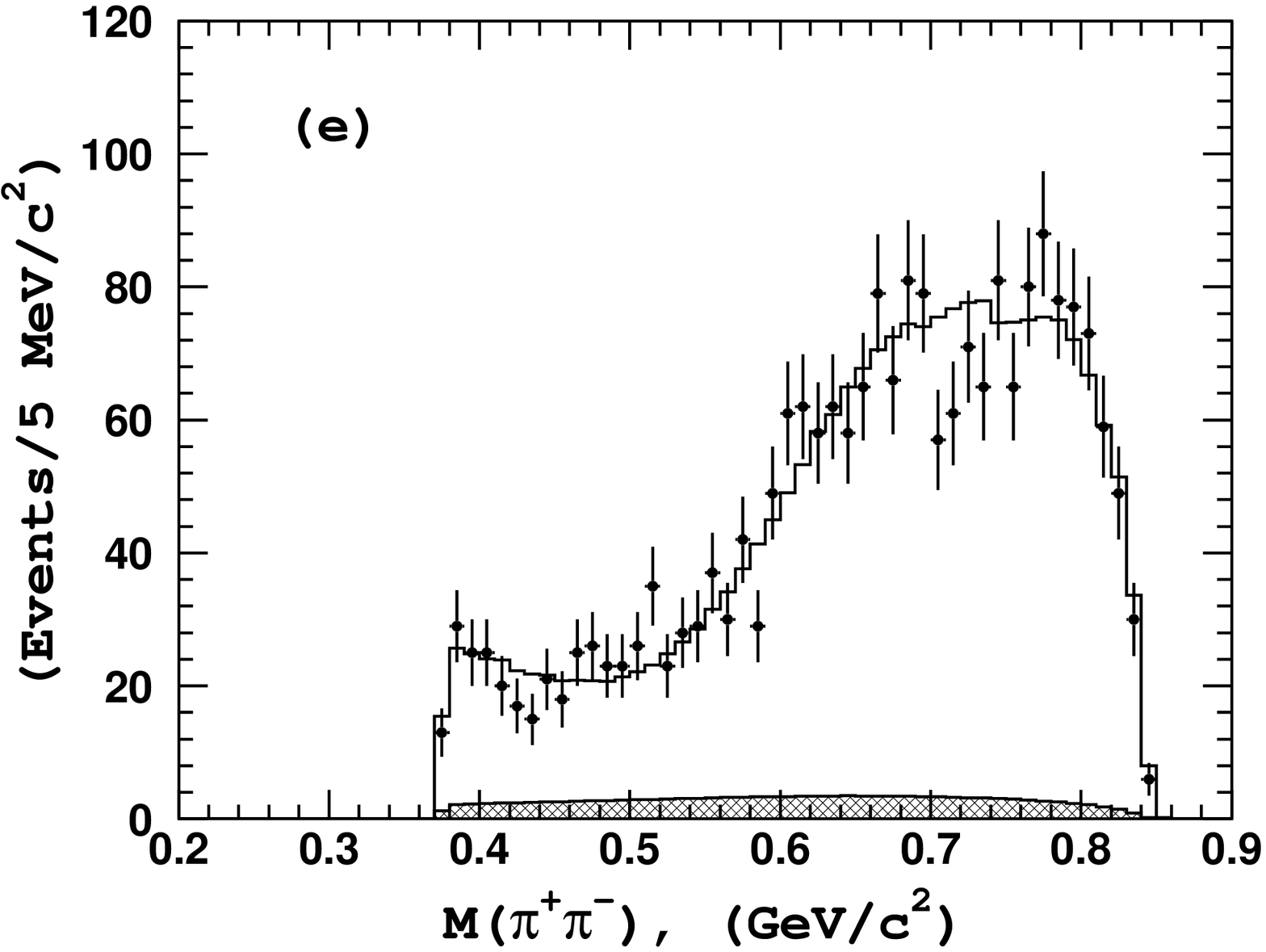} \hfill
  \includegraphics[width=0.32\textwidth]{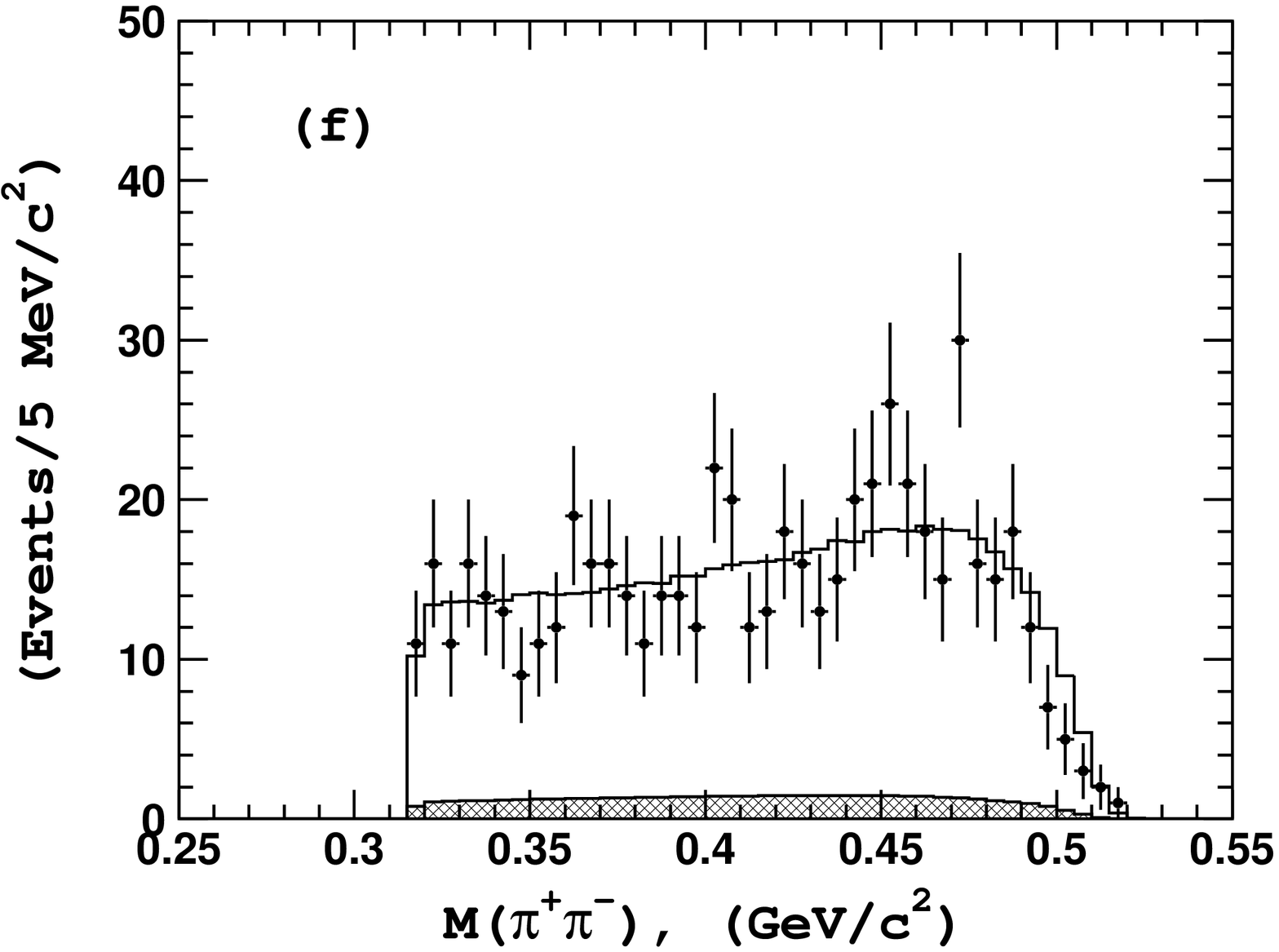} \\
  \caption{Comparison of fit results (open histogram) with experimental
           data (points with error bars) for $\Uo\pi^+\pi^-$ events 
           (left column), $\Ut\pi^+\pi^-$ events (middle column) and 
           $\Uh\pi^+\pi^-$ events (right column) in the signal region. 
           The hatched histograms show the background component.}
\label{fig:y3spp-f-hh}
\end{figure}

Following the suggestion given in 
Refs.\cite{Voloshin:2006ce,Voloshin:2007dx},
the non-resonant amplitude $A_{NR}$ is parameterized as 
\[
A_{\rm NR} = a\cdot e^{i\delta^{\rm nr}_1} +
             b\cdot e^{i\delta^{\rm nr}_2} \cdot s_3,
\]
where $s_3 = m^2(\pi^+\pi^-)$, $a_{\rm nr}$, $b_{\rm nr}$, $\delta^{\rm nr}_1$ 
and $\delta^{\rm nr}_2$ are free parameters of the fit ($s_3$ is not an 
independent variable and can be expressed via $s_1$ and $s_2$ but we prefer 
to keep it here for simplicity).

The logarithmic likelihood function ${\cal{L}}$ is then constructed as 
\[
{\cal{L}} = -2\sum{\log(f_{\rm sig}S(s_1,s_2) + (1-f_{\rm sig})B(s_1,s_2))},
\]
where $S(s_1,s_2)$ is formed from $|M(s_1,s_2)|^2$ convolved with the detector 
resolution and $f_{\rm sig}$ is the fraction of signal events in the data 
sample. Results of fits to $\Uf\to\Un\pp$ signal events are shown in 
Fig.~\ref{fig:y3spp-f-hh}, where one-dimensional projections of the data
and fits are presented. 

\begin{table}[!t]
\vspace*{1mm}
  \caption{Summary of results on fractions of individual quasi-two-body
channels contributiong to $\Uf\to\Un\pp$ three-body decays.}
  \medskip
  \label{tab:fracs}
\centering
  \begin{tabular}{lccc} \hline \hline
 Final state    &  ~~~~$\Uo\pp$~~~~    &
                   ~~~~$\Ut\pp$~~~~    &
                   ~~~~$\Uh\pp$~
\\ \hline
${\cal{B}}(Z^\mp_b(10610)\pi^\pm$)$\times
 {\cal{B}}(Z^\mp_b(10610)\to\Un\pi^\pm$), \% 
                       &  $2.54^{+0.86+0.13}_{-0.51-0.55}$
                       &  $19.6^{+3.5+1.9}_{-3.1-0.6}$     
                       &  $26.8^{+6.6}_{-3.9}\pm1.5$ \\
${\cal{B}}(Z^\mp_b(10650)\pi^\pm$)$\times
 {\cal{B}}(Z^\mp_b(10650)\to\Un\pi^\pm$), \% 
                       &  $1.04^{+0.65+0.07}_{-0.31-0.12}$
                       &  $5.77^{+1.44+0.27}_{-0.96-1.56}$ 
                       &  $11.0^{+4.2}_{-2.3}\pm0.7$ \\
${\cal{B}}(\Un f_2(1270)$)$\times
 {\cal{B}}(f_2(1270)\to\pp$), \%
                       &  $15.6\pm1.4\pm2.1$
                       &  $2.81^{+0.84+0.63}_{-0.56-0.86}$ 
                       &  $-$ \\
Total $S-$wave, \%     &  $89.2\pm3.0\pm2.4$
                       &  $105.6\pm4.1\pm2.6$              
                       &  $45.6\pm5.3\pm0.8$ \\
\hline \hline
\end{tabular}
\end{table}

Results on the $Z_b$ parameters are reported in Ref.~\cite{ypp}. 
Here, we report fractions of intermediate channels contributing to each
three-body final state. The results are summarized in 
Table~\ref{tab:ypp-frac}, where the central values are determined from fits
with the nominal model. Statistical uncertainties are determined from fits
to multiple toy MC samples generated according to the nominal model. This
allows us to account for correlations between various channels. In general, 
we find that all the three-body $\Uf\to\Un\pp$ decays are dominated by the 
$S$-wave channels with some statistically significant $D$-wave contribution. 

The dominant systematic uncertainty in the fractions of individual channels 
contributing to three-body decays comes from the model uncertainty. We 
estimate this uncertainty by fitting the signal with various modifications
of the nominal model. For example, we vary the parameterization of the 
non-resonant amplitude or replace the $\Un f_2(1270)$ amplitude with a 
$D$-wave non-resonant component.

% =============================================================================
% =============================================================================
% =============================================================================

\section{\boldmath Analysis of $\Uf\to\hm\pp$}

In the analysis of the $\Uf\to\hm\pp$ decays, we perform an inclusive 
reconstruction of signal events utilizing the recoil mass, $\mmpp$, 
associated with a $\pp$ pair. The selection requirements are identical to
those described in Ref.~\cite{hbp}. The continuum $\ee\to q\bar{q}$
($q=u,\;d,\;s,\;c$) background is suppressed by a requirement on the ratio
of the second to zeroth Fox-Wolfram moments $R_2<0.3$~\cite{Fox-Wolfram}.
We select $\pi^{\pm}$ candidates that originate from the vicinity of the 
interaction point and are positively identified as pions based on the CDC
($dE/dx$), TOF and ACC information. We reject tracks that are identified as
electrons. 

\begin{figure}[!tbp]
\includegraphics[width=0.48\textwidth]{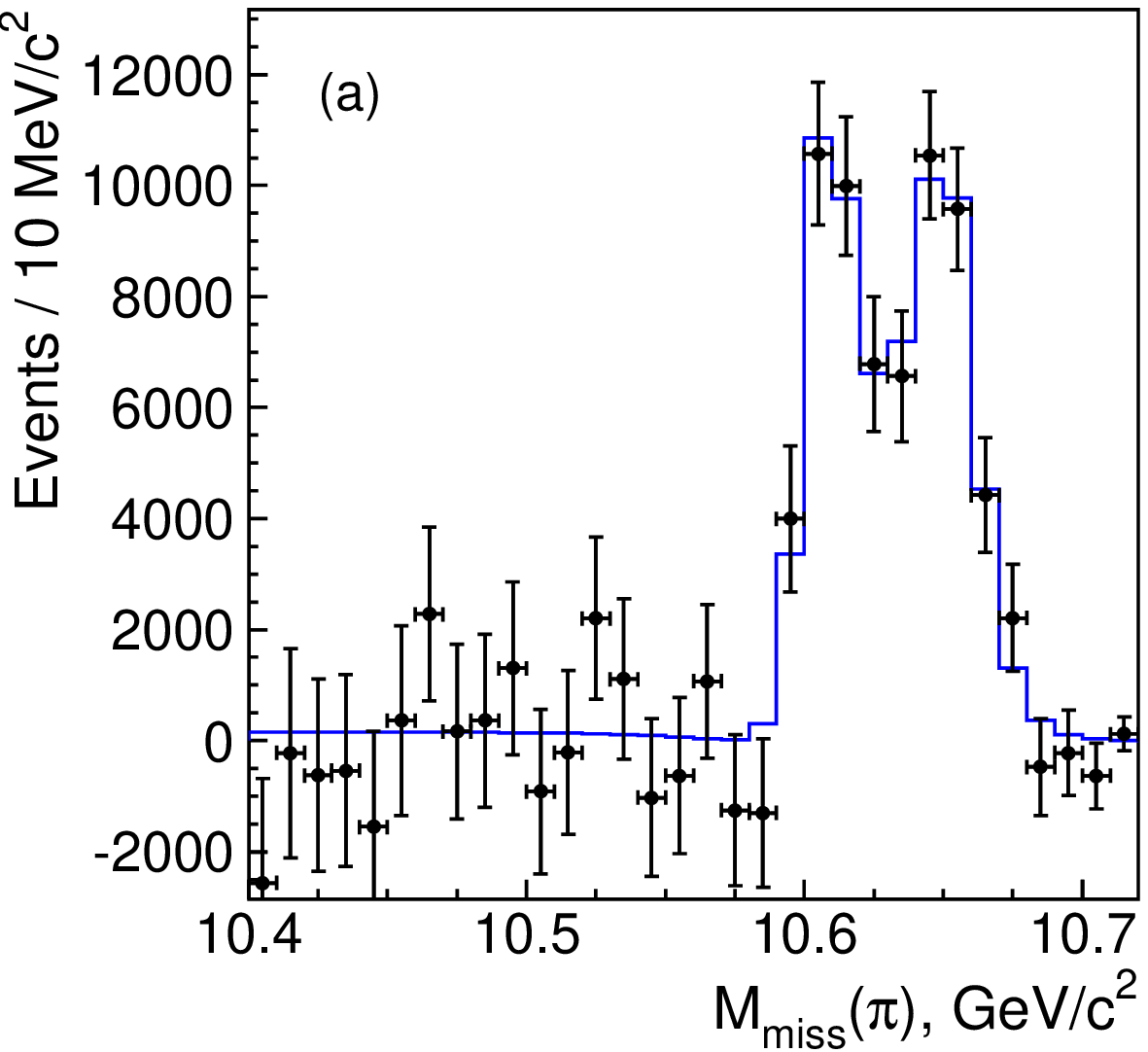} \hfill
\includegraphics[width=0.48\textwidth]{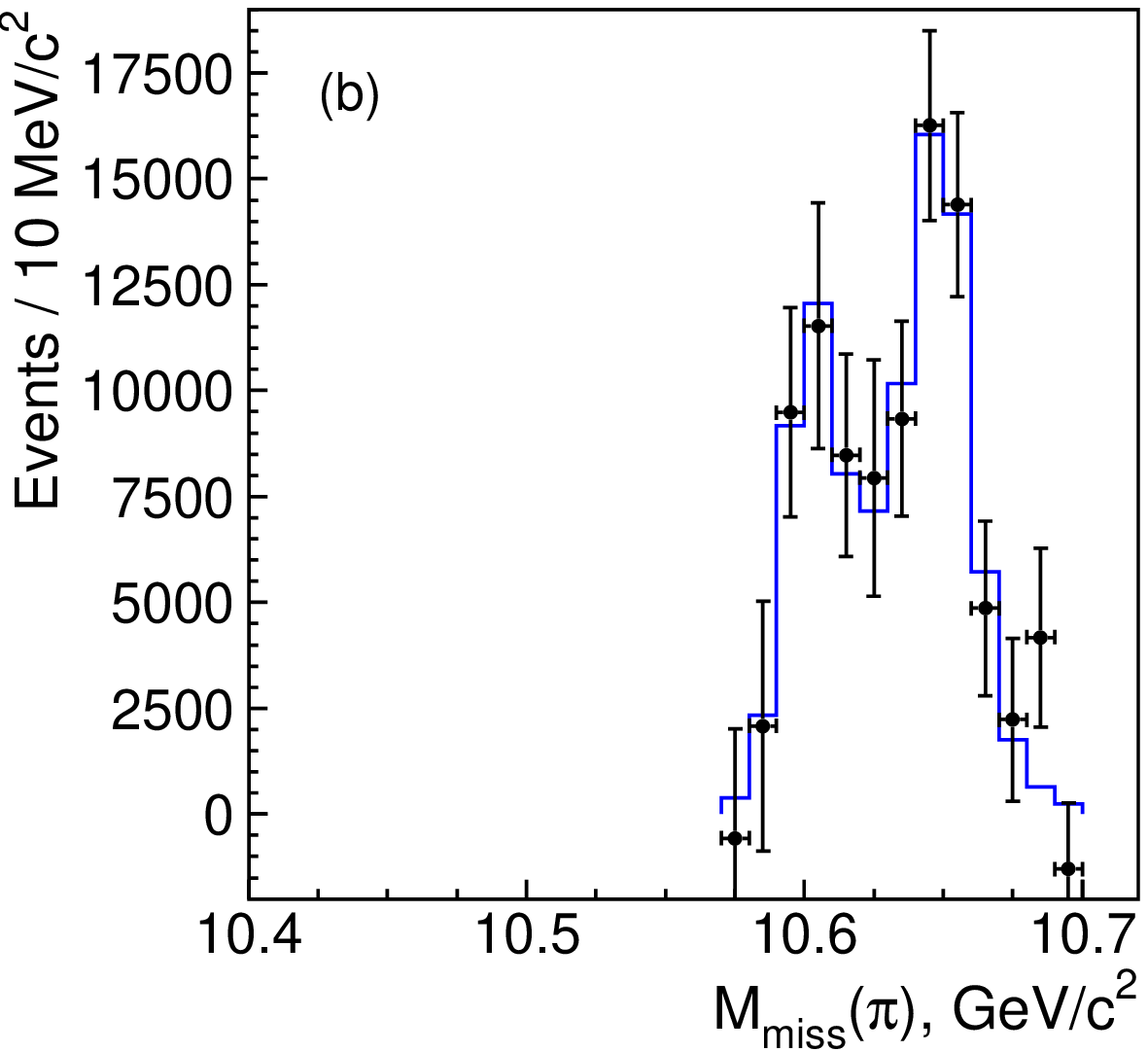}
\caption{ The yield (dots with error bars) of (a) $\hb$ and (b) $\hp$,
          as a function of $\mmp$. The solid histogram shows the results of
          the fit.}
\label{nhbn_vs_mmp}
\end{figure}

Because of the extremely high combinatorial background, a Dalitz analysis
of the $\Uf\to\hm\pp$ decay is challenging. Instead, we study the
one-dimensional projection by fitting the $\mmpp$ spectra in bins of the
$\hb\pi^{\pm}$ mass. We define the $\hb\pi^{\pm}$ mass as the recoil mass,
$M_r(\pi^{\mp})$, associated with a single charged pion. We symmetrize the
distributions by combining the $\mmpp$ spectra corresponding to $M_r(\pi^+)$
and $M_r(\pi^-)$ bins and restrict the analysis to the $\mmp>10.40\,\gevm$
($\mmp>10.57\,\gevm$) region for the $\hb\pp$ ($\hp\pp$) final state to
avoid double counting. Each $\mmpp$ spectrum is fit to extract the $\hb$
and the $\hp$ signal yields. The fitting function is a sum of a Crystal 
Ball function for the $\hm$ signal and a Chebyshev polynomial for the 
combinatorial background. We also account for the $\Ut$ signal and a 
reflection from $\Uh\to\Uo\pp$ decay. Details of this analysis can be 
found in Ref.~\cite{hbp}. The $\hm$ yields as a function of the 
$M_r(\pi^{\mp})$ are shown in Fig.~\ref{nhbn_vs_mmp}, where a clear two-peak
structure is apparent for both $\hb\pp$ and $\hp\pp$ final states.

We perform a $\chi^2$ fit of the $\mmp$ distributions to a coherent sum of
two $P$-wave Breit-Wigner amplitudes and a non-resonant contribution:
\begin{equation}
|BW_1(s,M_1,\Gamma_1)+ae^{i\phi}BW_1(s,M_2,\Gamma_2)
+be^{i\psi}|^2\frac{qp}{\sqrt{s}},
\label{hb_fit_fun}
\end{equation}
where $\sqrt{s}\equiv\mmp$; the variables $M_k$, $\Gamma_k$ ($k=1,2$),
$a$, $\phi$, $b$ and $\psi$ are free parameters; $\frac{qp}{\sqrt{s}}$
is a phase-space factor, where $p$ ($q$) is the momentum of the pion
originating from the $\Uf$ ($Z_b$) decay measured in the rest frame of
the corresponding mother particle.  The $P$-wave Breit-Wigner
amplitude is expressed as
$BW_1(s,M,\Gamma)=\frac{\sqrt{M\,\Gamma}\,F\,(q/q_0)}{M^2-s-iM\,\Gamma}$,
where $F$ is the $P$-wave Blatt-Weisskopf form factor
$F=\sqrt{\frac{1+(q_0R)^2}{1+(qR)^2}}$~\cite{blatt-weisskopf}, $q_0$
is the daughter momentum calculated using the pole mass of its mother,
$R=1.6\,\gev^{-1}$.  The function in Eq.~(\ref{hb_fit_fun}) is convolved 
with the detector resolution function ($\sigma=5.2\,\mevm$), integrated
over the histogram bin and corrected for the reconstruction efficiency.
The result of the fit is shown in Fig.~\ref{nhbn_vs_mmp}; fractions of
individual amplitudes are given in Table~\ref{tab:hbp-frac}. 

The non-resonant contribution in the $\hb\pp$ final state is found to be
consistent with zero, $b=0.18^{+0.22}_{-0.56}$ with the central value for
its fit fraction of only 3.2\% and an upper limit of 22\% at 95\% C.L.
This is in accordance with the expectation that the non-resonant amplitude
is suppressed due to the heavy quark spin flip. In the $\hb\pp$ final state,
we are not sensitive to the non-resonant amplitude due to the limited phase
space and thus fixed the non-resonant component at zero.

To estimate the systematic uncertainty we vary the order of the Chebyshev
polynomial in the fits to the $\mmpp$ spectra. To study the effect of 
finite $\mmp$ binning, we shift the binning by half of the bin size. To study
the model uncertainty in the fits to the $\mmp$ distributions, we remove (add)
the non-resonant contribution in the $\hb$ ($\hp$) case. We also vary the
$R$ parameter of the Blatt-Weisskopf form factor in the range of 
$0-5\,\gev^{-1}$ (default value is $1.6\,\gev^{-1}$) and find the associated
systematic effect to be negligible. In Ref.~\cite{hbp}, we find that the 
resolution in data could be larger than in MC by typically 10\%; we increase
the width of the resolution function by 10\% to account for possible 
difference between data and MC simulation. The maximum change of parameters
for each source is used as an estimate of its associated systematic error.
All systematic uncertainty contributions have been added in quadrature
to obtain the total systematic uncertainty.

\begin{table}[!t]
\caption{ Fit fractions of various components in three-body $\Uf\to\hm\pp$
signals. }
\label{tab:hbp-frac}
\renewcommand{\arraystretch}{1.1}
\begin{ruledtabular}
\begin{tabular}{l|cc}
& $\hb\pp$ & $\hp\pp$ \\
\hline
non-resonant   & 3.2\% ($<22$\% at 90\% C.L.) & -- \\
$\Zbl\pi^\pm$    & $(42.3^{+9.5}_{-12.7}\,^{+6.7}_{-0.8})\%$  
               & $(35.2^{+15.6}_{-9.4}\,^{+0.1}_{-13.4})\%$ \\
$\Zbh\pi^\pm$    & $(60.2^{+10.3}_{-21.1}\,^{+4.1}_{-3.8})\%$ 
               & $(64.8^{+15.2}_{-11.4}\,^{+6.7}_{-15.5})\%$ \\
\end{tabular}
\end{ruledtabular}
\end{table}

% =============================================================================
% =============================================================================
% =============================================================================

\section{Results}

To calculate branching fractions for the observed three-body 
$\UFS\to\bbstpi$ and $\UFS\to\bstbstpi$ signals, one needs to account for
the non-uniform distribution of signal events over the phase space. The
corrected efficiencies are found to be $12.25\pm0.06$\% and $11.0\pm0.1$\%,
for the $\UFS\to\bbstpi$ and $\UFS\to\bstbstpi$ modes, respectively. 
The three-body branching fractions are then calculated as
$$
  {\cal{B}}(\UFS\to\bbpi) = 
\frac{N_{\bbpi}}
{L\cdot\sigma(e^+e^-\to\UFS)\cdot 
 {\cal{B}}(B\to f)\cdot\varepsilon_{\bbpi}\cdot\alpha},
$$
where $L=121.4$~fb$^{-1}$ is the total integrated luminosity and
$\sigma({\ee\to\Uf})=0.340\pm0.016$~nb~\cite{y5scsec} is the cross section 
of $\ee$ annihilation into the $\Uf$ state at a c.m.\ energy of 10865~MeV.
Using world average results for the secondary branching fractions~\cite{PDG},
the combined fraction of $B$ meson decays to all reconstructed final states
including secondary branching fractions is found to be 
${\cal{B}}(B\to f)=(143\pm15)\times10^{-5}$ (neutral and charged $B$ combined). 
Finally, one also needs to correct for the oscillation of neutral $B$ mesons.
After time integration, the fraction of oscillated neutral $B$ mesons is 
equal to $f_{\rm osc}=0.19$. The correction factor $\alpha$ is calculated as:
$$
 \alpha = 
\frac{{\cal{B}}(B^+\to f^+)\cdot\varepsilon_{\bbpi} +
      {\cal{B}}(B^0\to f^0)\cdot\varepsilon_{\bbpi}(1-f_{\rm osc})}
{{\cal{B}}(B^+\to f^+)\cdot\varepsilon_{\bbpi}+
 {\cal{B}}(B^0\to f^0)\cdot\varepsilon_{\bbpi}},
$$
where ${\cal{B}}(B^+\to f^+)$ and ${\cal{B}}(B^0\to f^0)$ are the total 
fractions of $B$ decays to charged and neutral final states (including 
secondary fractions), respectively. From signal MC, one gets
$\alpha = 0.8978$. This results in 
${\cal{B}}(\UFS\to\bbstpi)=(28.3\pm2.9\pm4.6)\times10^{-3}$ and 
$\UFS\to\bstbstpi=(14.1\pm1.9\pm2.4)\times10^{-3}$. For the 
$\UFS\to BB\pi$ decay, we calculate a 90\% confidence level upper limit of
$\UFS\to BB\pi<4.0\times10^{-3}$ (including systematic uncertainty).

The dominant sources of systematic uncertainties for the three-body branching 
fractions are the uncertainties in the secondary branching fractions, the 
uncertainty in the reconstruction efficiency and in the signal yield 
extraction and the uncertainty in the $\sigma(e^+e^-\to \UFS)$ cross section.
The overall systematic uncertainties for the three-body branching fraction
are estimated to be $17.5$\%, $16.3$\% and $16.9$\% for the $BB\pi$, 
$\bbstpi$ and $\bstbstpi$ final states, respectively. 

\begin{table}[!t]
\vspace*{1mm}
  \caption{Results on three-body $\Uf\to\Un\pi^+\pi^-$ branching fractions.}
  \medskip
  \label{tab:ypp-frac}
\centering
  \begin{tabular}{lccc} \hline \hline
 Final state    &  ~~~~~~~$\Uo\pp$~~~~~~~    &
                   ~~~~~~~$\Ut\pp$~~~~~~~    &
                   ~~~~~~~$\Uh\pp$~~~
\\ \hline
Signal Yield    &  $2090\pm73$  &  $2476\pm97$  & $628\pm41$    \\
Efficiency, \%  &    $45.9$     &    $39.0$     &   $24.4$      \\
{\cal{B}}($\Un\to\mu^+\mu^-$)~\cite{PDG}, \% 
                & $2.48\pm0.05$ & $1.93\pm0.17$ & $2.18\pm0.21$ \\
{\cal{B}}($\Uf\to f$), $10^{-3}$ 
                & $4.45\pm0.16\pm0.35$  
                & $7.97\pm0.31\pm0.96$ 
                & $2.88\pm0.19\pm0.36$ \\ 
From Ref.~\cite{jack}, $10^{-3}$ 
                & $5.3\pm0.3\pm0.5$
                & $7.8\pm0.6\pm1.1$
                & $4.8^{+1.8}_{-1.5}\pm0.7$  \\
\hline \hline
\end{tabular}
\end{table}

The branching fractions of the three-body $\Uf\to\Un\pp$ decays are calculated
with the following formula:
\begin{equation}
 {\cal{B}}(\Uf\to\Un\pp) = \frac{N_{\Un\pp}}
{L\cdot\sigma({e^+e^-\to\Uf})\cdot {\cal{B}}(\Un\to\uu)\cdot\varepsilon_{\Un\pp}}.
\nonumber
\end{equation}
The reconstruction efficiencies (including trigger efficiency) 
$\varepsilon_{\Un\pp}$ are determined from the signal MC, $\Un\to\uu$ 
fractions~\cite{PDG}; the final results are given in Table~\ref{tab:ypp-frac}.
In determination of the reconstruction efficiencies, we use signal MC events
generated according to the results of the Dalitz fit with the nominal model.
The main systematic uncertainties in three-body fractions come from the
$\sigma({\ee\to\Uf})$ cross section (4.7\% for all channels), the 
$\Un\to\uu$ branching fractions (2.0\%, 8.8\% and 9.6\% for $n=1,2,3$,
respectively), the $\Un$ signal yield (4.5\%, 5.3\% and 4.9\% for $n=1,2,3$,
respectively), and the MC tracking efficiency of 4\% for all channels.
The overall systematic uncertainty is 7.9\%/12.0\%/12.4\% for 
$\Upsilon(1S/2S/3S)\pp$, respectively. The results for the $\Uf\to\Un\pp$ 
fractions are to be compared
%\footnote{Note, that in previous measurements $\sigma({\ee\to\Uf})=
%0.302\pm0.015$~nb have been used.} 
with previous measurements by Belle with
a data sample of 21~fb$^{-1}$~\cite{jack} (see Table~\ref{tab:ypp-frac}). 
We find the two sets of measurements to be consistent within uncertainties.

\begin{table}[!b]
\centering
\caption{List of branching fractions for the $Z^+_b(10610)$ and 
         $Z^+_b(10650)$ decays.}
\medskip
\label{tab:zfracs}
  \begin{tabular}{lcc}  \hline \hline
  ~Channel~\hspace*{80mm}  & \multicolumn{2}{c}{Fraction, \%}   \\
              & ~~~~~~~~~~~$\Zbl$~~~~~~~~~~~  & ~~$\Zbh$~~      \\
\hline 
 $\Upsilon(1S)\pi^+$      & $0.32\pm0.09$ & $0.24\pm0.07$      \\
 $\Upsilon(2S)\pi^+$      & $4.38\pm1.21$ & $2.40\pm0.63$      \\
 $\Upsilon(3S)\pi^+$      & $2.15\pm0.56$ & $1.64\pm0.40$      \\
 $h_b(1P)\pi^+$           & $2.81\pm1.10$ & $7.43\pm2.70$      \\
 $h_b(2P)\pi^+$           & $4.34\pm2.07$ & $14.8\pm6.22$      \\
 $B^+\bar{B}^{*0}+\bar{B}^0B^{*+}$
                          & $86.0\pm3.6$  &    $-$             \\
 $B^{*+}\bar{B}^{*0}$     &    $-$        & $73.4\pm7.0$       \\
\hline \hline
  \end{tabular}
\end{table}

Using results of the fit to the $M_r(\pi)$ spectra with the nominal model 
(see Table~\ref{tab:results}) and the results of the analysis of the 
$\UFS\to\Upsilon(nS)$, $n=1,2,3$ and $\UFS\to h_b(mP)$, $m=1,2$ 
decays, one can measure the ratio of the branching fractions:
\begin{eqnarray}
 \frac{{\cal{B}}(\Zbl\to BB^*)}{\sum_n{{\cal{B}}(\Zbl\to\Upsilon(nS)\pi)}+
                          \sum_m{\Zbl\to h_b(mP)}} =
6.2\pm0.7\pm1.3^{+0.0}_{-1.8} \nonumber
\label{eq:zblfrac}
\end{eqnarray}
and
\begin{equation}
\frac{{\cal{B}}(\Zbh\to B^*B^*)}{\sum_n{{\cal{B}}(\Zbh\to\Upsilon(nS)\pi)}+
                          \sum_m{\Zbh\to h_b(mP)}} = 
2.8\pm0.4\pm0.6^{+0.0}_{-0.4}. \nonumber
\label{eq:zbhfrac}
\end{equation}

We also find it useful to calculate the relative fractions for $Z_b$
decays assuming that thy are saturated by the already observed
$\Upsilon(nS)$ $(n=1,2,3)$, $h_b(mP)$ $(m=1,2)$, and $B^*B^{(*)}$
channels. The results are summarized in Table~\ref{tab:zfracs}. We do not
include the $\Zbh\to BB^*$ channel in the table as this decay mode has
marginal significance. However, if the central value is used, its fraction 
would be $25.4\pm10.2$\%. All other fractions would be reduced by a factor 
of 1.33.

\section*{Acknowledgments}

We thank the KEKB group for the excellent operation of the accelerator; the 
KEK cryogenics group for the efficient operation of the solenoid; and the 
KEK computer group, the National Institute of Informatics, and the PNNL/EMSL
computing group for valuable computing and SINET4 network support. 
We acknowledge support from the Ministry of Education, Culture, Sports, 
Science, and Technology (MEXT) of Japan, the Japan Society for the Promotion
of Science (JSPS), and the Tau-Lepton Physics Research Center of Nagoya
University; the Australian Research Council and the Australian Department of
Industry, Innovation, Science and Research; the National Natural Science 
Foundation of China under contract No.\, 10575109, 10775142, 10875115 and 
10825524; the Ministry of Education, Youth and Sports of the Czech Republic
under contract No. LA10033 and MSM0021620859; the Department of Science and
Technology of India; the Istituto Nazionale di Fisica Nucleare
of Italy; the BK21 and WCU program of the Ministry Education Science and 
Technology, National Research Foundation of Korea, and GSDC of the Korea 
Institute of Science and Technology Information; the Polish Ministry of 
Science and Higher Education; 
%the Ministry of Education and Science of the 
%Russian Federation and the Russian Federal Agency for Atomic Energy;
the Ministry of Education and Science of the Russian
Federation, the Russian Federal Agency for Atomic Energy and the
Russian Foundation for Basic Research Grant RFBR 12-02-01296;
the Slovenian Research Agency; the Swiss National Science Foundation; the
National Science Council and the Ministry of Education of Taiwan; and the 
U.S. Department of Energy and the National Science Foundation. This work 
is supported by a Grant-in-Aidfrom MEXT for Science Research in a Priority 
Area (``New Development of Flavor Physics''), and from JSPS for Creative 
Scientific Research (``Evolution of Tau-lepton Physics'').

%%%%%%%%%%%%%%%%%%%%%%%%%%%%%%%%%%%%%%%%%%%%%%%%%%%%%%%%%%%%%%%%%%%%%%%%%%%%%%%
%%%%%%%%%%%%%%%%%%%%%%%%%%%%%%%%%%%%%%%%%%%%%%%%%%%%%%%%%%%%%%%%%%%%%%%%%%%%%%%
%%%%%%%%%%%%%%%%%%%%%%%%%%%%%%%%%%%%%%%%%%%%%%%%%%%%%%%%%%%%%%%%%%%%%%%%%%%%%%%

\end{document}